\documentclass[a4paper,11pt]{article}
\usepackage{slashed}
\usepackage[affil-sl,auth-sc]{authblk}
\usepackage{amssymb,amsmath,amsbsy}
\usepackage{xfrac}
\usepackage{mathrsfs}
\usepackage{mathbbol}
\usepackage{bm}
\usepackage{appendix}
\usepackage{hyperref}
\usepackage{bookmark}
\usepackage[top=1.2in, bottom=1.2in,left=0.9in,includefoot]{geometry}               
\usepackage{graphicx}
\usepackage{cite}
\usepackage{float}
\usepackage{caption}
\usepackage{wasysym}
\usepackage{amsthm}
\usepackage{romannum}
\newif\ifContLineOne
\newif\ifContLineTwo
\newif\ifContLineThree
\def\conC#1{\vbox{\ialign{##\crcr
  \ifContLineThree\hrulefill\else\vphantom{\hrulefill}\fi\crcr
  \noalign{\kern3.2pt\nointerlineskip}
  \ifContLineTwo\hrulefill\else\vphantom{\hrulefill}\fi\crcr
  \noalign{\kern3.2pt\nointerlineskip}
  \ifContLineOne\hrulefill\else\vphantom{\hrulefill}\fi\crcr
  \noalign{\nointerlineskip}
  $\hfil\textstyle{\vbox to 14pt{}#1}\hfil$\crcr}}}
\def\DrawLeg#1#2{
  \kern-.2pt              
  \dimen2 =#1             
  \advance\dimen2 by 2pt  
  \dimen3 = 10.6pt        
  \dimen4 =3.6pt          
  \advance\dimen3 by -\dimen2 
  \multiply\dimen4 by #2
  \advance\dimen3 by \dimen4
  \raise\dimen2 \hbox{\vrule height\dimen3 width .4pt} 
  \kern-.2pt}             
\def\begC#1#2{\setbox0 =\hbox{$\textstyle{#2}$}
  \dimen0=.5\wd0 \dimen1=\ht0
  \conC{\hskip\dimen0}
  \count255=#1
  \ifnum\count255 =1 \ContLineOnetrue\else
  \ifnum\count255 =2 \ContLineTwotrue\else
  \ifnum\count255 =3 \ContLineThreetrue\fi\fi\fi
  \DrawLeg{\dimen1}{\count255}
  \conC{\hskip\dimen0}
  \kern-\dimen0\kern-\dimen0 \box0}
\def\endC#1#2{\setbox0 =\hbox{$\textstyle{#2}$}
  \dimen0=.5\wd0 \dimen1=\ht0
  \conC{\hskip\dimen0}
  \count255=#1
  \ifnum\count255 =1 \ContLineOnefalse\else
  \ifnum\count255 =2 \ContLineTwofalse\else
  \ifnum\count255 =3 \ContLineThreefalse\fi\fi\fi
  \DrawLeg{\dimen1}{\count255}
  \conC{\hskip\dimen0}
  \kern-\dimen0\kern-\dimen0 \box0}

\def\theequation{\arabic{section}.\arabic{equation}}
\def\beq {\begin{equation}}
\def\eeq {\end{equation}}
\def\bea {\begin{eqnarray}}
\def\eea {\end{eqnarray}}

\newcommand{\br}{\begin{eqnarray}}
\newcommand{\er}{\end{eqnarray}}
\newcommand{\be}{\begin{equation}}
\newcommand{\ee}{\end{equation}}

\newcommand{\lsim}{\raisebox{-0.13cm}{~\shortstack{$<$ \\[-0.07cm] $\sim$}}~}
\newcommand{\gsim}{\raisebox{-0.13cm}{~\shortstack{$>$ \\[-0.07cm] $\sim$}}~}

\newcommand{\nn}{\nonumber}

\def\invfb{fb^{-1}}
\def\hsm{H_{SM}}
\def\mh{\rm m_{H_1}}
\def\ma{\rm m_{A_1}}
\def\mhh{\rm m_{H_2}}

\def\hu{\rm {H_u}}
\def\hd{\rm {H_d}}
\def\t1{\widetilde {t_1}}
\def\N0{\widetilde \chi}
\def\mst1{m_{\t1}}
\def\103{\times 10^3}

\usepackage{color}
\hypersetup{linktocpage=true}
\usepackage[all]{hypcap}
\title{\LARGE\bf Light Higgs Bosons in NMSSM at the LHC }
\author[1]{Monoranjan Guchait\thanks{guchait@.tifr.res.in}}
\author[2]{Jacky Kumar\thanks{jka@tifr.res.in}}
\affil[1,2]{\footnotesize Department of High Energy Physics,\ Tata Institute of Fundamental Research, 400 005, Mumbai, India }

\def\invfb{fb^{-1}}
\def\hsm{H_{SM}}
\def\mh{m_{H_1}}
\def\ma{m_{A_1}}
\def\mhh{m_{H_2}}

\def\t1 {\widetilde {t_1}}
\def\N1{\widetilde \chi_1^0}
\def\N2{\widetilde \chi_2^0}
\def\N3{\widetilde \chi_3^0}
\def\N0{\widetilde \chi^0}
\def\C1{\widetilde \chi_1^{\pm}}
\def\mst1 {m_{\t1}}
\def\br {\begin{eqnarray}}
\def\er {\end{eqnarray}}

\date{}
\begin{document}
\maketitle
\begin{abstract}
{\noindent\normalsize
The next-to-minimal supersymmetric standard model (NMSSM)  
with an extended Higgs sector {{offers at least one}} Higgs boson as the
Standard model (SM) like Higgs with a mass around 125 GeV. 
In this work, we revisit the mass spectrum and couplings of non-SM-like 
Higgs bosons taking into consideration {{most relevant}} constraints 
and identify 
the relevant parameter space. The discovery potential
of these non-SM-like Higgs bosons, apart from their masses,  
is guided by their couplings with gauge bosons and fermions which are 
very much parameter space sensitive. We evaluate the rates of 
productions of these non-SM-like Higgs bosons at the LHC for a 
variety of decay channels in the allowed region of the parameter space. 
Although $b \bar b, \tau \tau$ modes appear to be the most 
promising {{decay channels}}, but for a substantial region of 
parameter space the two-photon decay mode has a remarkably  
large rate. In this study we emphasize that this diphoton mode can be  
exploited to find the {{non-SM-like Higgs bosons of the NMSSM}} 
and can also be a   
potential avenue to distinguish the NMSSM from the MSSM. In addition, 
we discuss briefly the various detectable signals of these non SM 
Higgs bosons at the LHC.}    
\end{abstract}
\vskip .5 true cm
\newpage
\section{Introduction}
\label{intro}
The recent discovery of the Higgs 
particle~\cite{Chatrchyan:2012ufa,Aad:2012tfa} 
has completed the particle family in
the Standard model (SM), although more precise measurements are required to
establish this claim strongly.
On the other hand, there are various new physics models which can accommodate 
this Higgs candidate and are found to be consistent with the current  
measurements of Higgs couplings strength
~\cite{Khachatryan:2014jba,Barreiro:2014wwa,Aad:2015zhl}. For example,
the phenomenological models based on supersymmetric (SUSY) theory 
can interpret this {{Higgs candidate as their lightest 
Higgs boson}} \cite{Hall:2011aa,Arbey:2011ab}. However, it is to be noted 
that to accommodate a Higgs boson of mass $\rm \sim$125~GeV
in the theory of Minimal Supersymmetric Standard Model(MSSM), 
one requires a substantial contribution to the Higgs mass from higher 
order correction which is very close to 
the tree level value of it. Of course, this can be 
achieved in a very suitable region of SUSY parameter space, 
in particular with the contribution from the 
third generation squark sector along with some 
degrees of fine-tuning~\cite{Barbieri:1987fn,Kitano:2006gv}.

The next-to-minimal supersymmetric standard model (NMSSM), an extension 
of MSSM with an additional Higgs singlet was proposed 
to solve the $\rm \mu$ problem
\cite{Ellis:1988er,Drees:1988fc}.
Interestingly, it has been found that this model can provide 
a SM-like Higgs boson of mass around 125 GeV more naturally (For example, see
Refs.~\cite{Barger:2006dh,Ellwanger:2009dp,Ellwanger:2011sk,King:2012is,
Vasquez:2012hn,
King:2012tr,Christensen:2013dra,Cao:2013gba,Kang:2012sy}
and references therein), and 
as a result, after the discovery of the 
Higgs particle~\cite{Chatrchyan:2012ufa,Aad:2012tfa}, it has received a 
lot of attention. In this model the 
additional interaction term between 
the singlet and the doublet Higgs superfields contributes substantially
to the Higgs mass, even at the tree level and thus reduces the required fine 
tuning significantly. 
The phenomenology of the NMSSM, more precisely of its 
Higgs sector, is quite rich due to the presence of this 
extra Higgs singlet and the corresponding interaction 
with the Higgs doublet. In the literature, there exist several studies 
carried out to study the NMSSM Higgs sector in 
the context of the LHC. 
The recent observation of the Higgs particle at the LHC and related 
measurements of its properties renewed the interest in the NMSSM leading
to more focused and dedicated activities. The implications of the 
current observation of the Higgs 
boson at the LHC and the measurements of its properties have been extensively
discussed in the context of the NMSSM by many authors. 
In the LHC experiment looking for the NMSSM Higgs signal in the 
current Run I data sets is one of the ongoing studies. 
In a very recent analysis, very light Higgs boson in the NMSSM 
was studied and from the non-observation of any signal, a bound on 
the Higgs production cross section times the corresponding 
Higgs branching ratio {{(BR)}} is presented~\cite{Chatrchyan:2012am,Aad:2015oqa}.

The NMSSM contains seven Higgs bosons, three are CP even Higgs 
scalars (${\rm H_i, i=1,2,3}$) and two are CP odd states, 
while there are two charged Higgs bosons.
In this work we revisit the Higgs sector of the NMSSM by studying 
the interplay of model parameters and Higgs masses and 
couplings. We explore the possible implications 
in the NMSSM Higgs sector of recent Higgs discovery 
by scanning the model parameter space for a wide range requiring one of 
the CP even Higgs bosons SM-like Higgs with mass around 125 GeV.
Performing a very comprehensive analysis taking into account various 
experimental and theoretical constraints, 
we delineate the region 
of parameter space which provides one of
the CP even Higgs boson SM-like. Corresponding to this 
allowed parameter space, we then discuss the phenomenology of, particularly,
light neutral non-SM-like CP even and odd Higgs bosons 
at the LHC.
It is observed that there exist a variety of decay channels of 
these light neutral non SM Higgs bosons with very 
reasonable {{BR}} depending on the parameters points. 
In this work, we study very systematically in detail  
the sensitivity of these BRs 
on model parameters and their impact on light non-SM-like Higgs boson
signals at the LHC. 
Moreover, in order to understand the feasibility of non-SM-like Higgs 
searches at the LHC, the rates of production of Higgs 
bosons in  
various decay channels are estimated for the center of mass 
energy $\rm \sqrt{s}=13$ TeV,
in the context of the interesting region
of parameter space.   
In fact due to many parameters dependence of the cross sections and 
BRs of the NMSSM Higgs bosons, it is difficult to present systematically
the variation of non SM Higgs boson production rates 
in various decay channels.
Hence, we follow the strategy to present the ranges of these rates 
by computing 
the minimum and maximum values in various decay channels for a given 
Higgs boson mass and the 
parameter space.
Naively, this estimation of rates provides hints about the detectability
of non-SM-like Higgs bosons at the LHC although detail understanding 
of the background level is required to make a final conclusion.
Interestingly, it is observed that the BR of non-SM-like Higgs in the  
di-photon channel is substantially large
for a certain region of parameter 
space~\cite{Christensen:2013dra,Ellwanger:2010nf}. 
Undoubtedly, this two photons channel appears as a striking feature of 
the NMSSM model. 
Perhaps, the di-photon channel can provide a robust 
mode to distinguish the NMSSM from the MSSM because of the fact that 
photon is a clean object to probe experimentally. In addition there are 
also other hadronic decay channels of non SM Higgs bosons which
can be useful as well in looking for Higgs signal in the NMSSM. 
 It is to be noted that the similar type of study is carried out,   
in particular in Ref.~\cite{King:2014xwa} where 
the main focus is to study only the pair production of Higgs. In this paper, 
we study the production of Higgs via all dominant channels 
focusing the range of masses for all Higgs bosons below 125 GeV,
the interesting range in the context of the present LHC experiments.
In order to discuss the Higgs signal we have very systematically 
discussed the BR of various decay channels of non SM-like Higgs bosons
including the two interesting channels, $gg$ and $c\bar c$. 
More importantly, in this current study we obtain rates for various 
decay channels estimating the Higgs boson production cross section 
following the model based calculation implemented in 
SuSHi~\cite{Liebler:2015bka}.

The paper is organized as follows. In section 2 we discuss very briefly 
the relevant parts of the NMSSM Higgs sector. In section 3, 
describing inputs and constraints used 
while scanning parameters, we discuss  
the allowed ranges of the masses, the couplings and decay BRs  
of non-SM-like light Higgs bosons.
Presenting the rates of the non-SM-like Higgs boson 
production in various channels in section 4, we discuss various
Higgs signals in section 5. Finally we summarize our results in section 6.

\section{NMSSM Higgs sector }
\label{nugm}
As mentioned in the introduction, the NMSSM Higgs sector contains one 
extra gauge singlet Higgs superfield $ \rm{\hat{ S}}$ in addition 
to two Higgs doublets ($\rm{ \hat{ H_u}}, \rm{\hat{ H_d}}$)
~\cite{Ellis:1988er,Drees:1988fc,Barger:2006dh} making the super potential as,
\br
{\rm W_{NMSSM}} = {\rm W_{MSSM}} + \lambda \hat S  \hat{\hu} 
\hat{\hd} + \frac{1}{3}\kappa \hat {\bf{\rm{S^3}}}, 
\label{eq:poteq}
\er
where $\rm{W_{MSSM}}$ corresponds to the super potential in MSSM with only 
two Higgs doublets without a $\mu$ term, 
$\rm \lambda$ and $\rm \kappa$ are dimensionless couplings. 
Recall that the primary goal in constructing the NMSSM is to generate 
the $\rm \mu$ term dynamically to ensure its value around the EW scale, and
thus solving the $\rm \mu$ problem. The last term in Eq.~\ref{eq:poteq} with 
$\rm{\hat S^3}$ is introduced to avoid Peccei-Quinn(PQ) 
symmetry~\cite{Peccei:1977hh}. Notice that this expansion 
respects explicit $Z_3$ symmetry. 
 {In the MSSM, $\rm W_{MSSM}$ depends 
on  $\hat{ \rm \hu}$ and $\hat{\rm  \hd}$, the Higgs super fields 
which couple with up-type quarks and down type  quarks and charged 
leptons respectively. }

Correspondingly, the soft-SUSY breaking potential for NMSSM Higgs sector 
is given by, 
\br
\rm V_{soft} =m^{2}_{\hu}|\hu|^{2} + m^{2}_{\hd}|\hd|^{2}  
+m^{2}_{S} |S|^{2} +\left [ \lambda A_{\lambda }  S \hu \hd +
\frac{1}{3} \kappa A_{\kappa} S^{3}+h.c\right]. 
\er
Here $\rm m_{\hu}$, $\rm m_{\hd}$, $\rm m_S$, $\rm A_\lambda$ and 
$A_\kappa$ are the soft breaking parameters. Notably, the parameters 
$\rm \lambda$, 
$\rm \kappa$, $\rm A_\lambda$ are taken to be positive.
The vacuum expectation value (VEV) of the singlet field, 
$\left <\rm{ S} \right >$=${v_s}$, 
in turn generates the effective $\mu$ term, 
$\rm{\mu_{eff}}$=$ \rm{\lambda v_s}$ and then  
restricting it at the order of EW scale.   
The neutral components of other Higgs doublets, 
get VEVs as $\left <\rm{ \hu^0} \right > = \rm{v_u}$ and
$\rm {\left <\hd^0 \right> =\rm v_d}$ and the constraint  
$\rm {m_W^2= {\frac{g^2}{4}(v_u ^2 + v_d^2)}} $ makes one of the VEVs 
as a free parameter and it is parametrized as 
$\rm{\tan\beta={\rm {v_u}}/{\rm{v_d}}}$.   

At the tree level, the Higgs sector of NMSSM 
have 9 parameters i.e, 
\br
\mathrm{\lambda,~ \kappa,~ \tan{\beta},~\mu_{eff},~ A_{\lambda},
~A_{\kappa}},\ \ m^{2}_{H_{u}},\ \ m^{2}_{H_{d}},\ \ m^{2}_{S}.  
\label{eq:para}
\er
{{But}} the minimization conditions of the scalar potential  
with respect to VEVs $v_{u},v_{d}$ and $v_s$ reduce another three parameters,
and hence, the NMSSM Higgs sector at the tree level is described 
by six independent parameters,
\br
\mathrm{\lambda,~ \kappa,~ \tan{\beta},~\mu_{eff},~ A_{\lambda},
~A_{\kappa}},
\label{eq:nmssmparam}
\er
unlike the MSSM  where only two parameters $m_A$, the mass of the 
pseudoscalar Higgs and $\rm{\tan\beta}$ are required. 
More detail discussions and 
review on the NMSSM Higgs sector can be found
in Refs.~\cite{Ellis:1988er,Drees:1988fc,Barger:2006dh,Ellwanger:2009dp,
Franke:1995tc, Miller:2003ay}.

The NMSSM Higgs sector contains 10 scalar degrees 
of freedom, out of which three are absorbed to give the masses 
of three gauge bosons, $\rm{ W^\pm}$ and Z leading to seven physical 
Higgs states.
Expansion of Higgs fields around the three
vevs $\rm{v_u, v_d}$ and $\rm{v_s}$ yield various Higgs mixing terms in 
the Lagrangian of which real and imaginary parts 
constitute 
{{two 3$\times$3 Higgs mass matrices corresponding to the CP-even and 
CP-odd scalars}}.
The diagonalization of these mass matrices by orthogonal matrices provides   
masses of three physical Higgs bosons. 
For instance, the diagonalization of the 3$\times$3 real mass matrix results
in the masses of three CP even Higgs bosons ($\rm H_1, H_2, H_3$) and     
the corresponding diagonalizing matrix relates the weak Higgs boson states, 
$\rm{ H_j^w  \equiv (\hu^R, \hd^R, S^R)}$ with the physical ones as,
\br
\rm{ H_i=S_{ij} H_j^{w}};\ \ \rm i,j=1,2,3, 
\er
where $\rm S_{ij}$ is the orthogonal matrix diagonalizing 
the $3 \times 3$ CP even Higgs mass 
matrix. Here the CP even mass eigenstates ($\rm{ H_i}$) arranged in 
increasing order of masses 
i.e $\rm{m_{H_1} < m_{H_2} < m_{H_3}} $. 
Consequently, the physical Higgs boson states turn out to be admixtures of 
both doublet ($\rm \hu,\rm \hd$) and singlet (S) states and naturally the 
corresponding 
mixing angles affect their couplings to fermions and gauge bosons.      
Similarly, the imaginary parts of Higgs fields form a $3\times 3$ mass 
matrix in the basis  $\rm{A_j^w \equiv (H_{u}^I, H_{d}^I, S^{I})}$ 
and is being  
diagonalized by an orthogonal matrix as, 
\br
\rm{A_i = P_{ij} A_j^w}, 
\er 
$\rm{A_i}$ stands for three CP odd states, 
$\rm{(A_1, A_2, G^0}$), where $\rm{G^0}$ is the Goldstone boson.
Defining a suitable 2$\times$2 matrix, the Goldstone boson 
state $\rm{G^0}$ can be rotated away leading only to two physical 
states $\rm{A_1}$ and $\rm{A_2}$ which are admixtures of doublet
and singlet states 
determined by the components of the orthogonal matrix $\rm{P_{ij}}$.   
The mixing angles i.e the matrix elements of the orthogonal 
matrices depend on the model parameters given by Eq.\ref{eq:nmssmparam}
~\cite{Ellwanger:2009dp,Franke:1995tc,Miller:2003ay} and they 
play an important role in determining the couplings 
of the Higgs bosons with fermions and gauge bosons making those 
very much parameter space sensitive.
For example, the couplings of CP even Higgs bosons 
$\rm{H_i}$ to fermions, of
both top and bottom type, are given by
~\cite{Ellwanger:2009dp,Franke:1995tc,Miller:2003ay}, 

\br
\rm g_{H_i t\bar t} :- \frac{m_t}{{\sqrt{2}v}\sin\beta} S_{i1}; \nn \\  
\rm g_{H_ib\bar b}: \frac{m_b}{{\sqrt{2}v}\cos\beta} S_{i2} ;\nn \\ 
\rm g_{H_i\tau\tau}: \frac{m_\tau}{{\sqrt{2}v}\cos\beta} S_{i2},
\label{eq:hcoup}
\er
and similarly for CP odd scalars, $\rm{A_i}$,
\br
\rm g_{A_it\bar t}: - { i} \frac{m_t}{{\sqrt{2} v}\sin\beta} P_{i2}; \nn \\ 
\rm g_{A_ib\bar b}: {i} \frac{m_b}{{\sqrt{2} v}\cos\beta} P_{i1}; \nn \\ 
\rm g_{A_i\tau\tau}: {i} \frac{m_\tau}{{\sqrt{2}v}\cos\beta} P_{i1}. 
\label{eq:acoup}
\er
Here $\rm{S_{i1}, S_{i2}}, \rm{(P_{i1} ,P_{i2} )}$ determine the 
amount of doublet composition 
in physical CP even(odd) Higgs boson states. 
Thus due to the presence of mixing angles, Higgs couplings either get 
suppressed or enhanced. For instance, if any of the Higgs state 
($\rm{H_i,A_i}$) 
be dominated by singlet components  
i.e $\rm{S_{i1}, S_{i2} \sim 0}$, $\rm{P_{i1},P_{i2} \sim 0}$ 
then the corresponding Higgs boson 
couplings with fermions and gauge bosons are severely suppressed.  
The ratio between the Higgs couplings given in 
Eq.~\ref{eq:hcoup}, \ref{eq:acoup} and its SM counter part is just a 
scale factor, called reduced coupling depending on mixing angle 
and $\rm{\tan \beta}$, e.g $\rm{C_u=\rm S_{i1}/\sin\beta}$ is the 
reduced coupling of $\rm{H_i}$ 
with up-type quark. Moreover, the orthogonality property of the mixing 
matrices(S,P) predicts a sharing of couplings among the corresponding 
Higgs boson states. For example if any of the CP even Higgs boson be SM 
like i.e $\rm{S_{i1}}$ or $\rm{S_{i2} \sim 1}$, then the corresponding 
couplings of other CP even Higgs bosons with fermions are heavily 
suppressed. The coupling of CP even Higgs bosons with gauge 
bosons (V=W,Z) are given by~\cite{Ellwanger:2009dp,Miller:2003ay}, 
\br
\rm{g_{H_i VV} = g^{SM}_{H_i,VV} \times \xi_i; \ \ \xi_i = \cos\beta S_{i2}+
\sin\beta S_{i1}},
\label{eq:sumrule1}
\er 
and because of orthogonality property of matrix S, one can show that 
the mixing angle follows the relation, 
\br
\rm{\sum \xi_i^2 =1}
\label{eq:sumrule2}.
\er 
As stated earlier this sum rule implies that if any one of the Higgs 
boson couples 
dominantly with a gauge boson, then the same couplings of other 
Higgs bosons are severely disfavoured.
The mixing angle $\rm{\xi_i}$ is constrained from the non observation 
of NMSSM Higgs bosons  
in LEP experiment~\cite{Schael:2006cr}.

In NMSSM the upper bound of lightest Higgs boson mass at tree level
is well known to be 
as~\cite{Ellis:1988er,Drees:1988fc,Franke:1995tc,Ellwanger:1993xa,
King:1995vk}, 
\br
\rm{m_{H_1}^2 \lsim M_Z^2 \cos^2 2\beta + \lambda^2 {v^2}\sin^22\beta},
\er
where the first term is similar to the upper bound(=$\rm M_Z\cos2\beta$) 
in MSSM and the second term originates due to the interaction 
between singlet and doublet Higgs field via $\rm{\lambda}$ term, 
in Eq.~\ref{eq:poteq}. 
Remarkably, the contribution of the second term pushes this 
upper bound substantially much above $\rm{M_Z \cos 2\beta}$, of course,
depending on the values of $\rm{\lambda}$ and $\rm{\tan\beta}$. 
Clearly, high (low) value of $\rm{\lambda}$ ($\rm{\tan\beta}$) is 
preferred to obtain a large value of the 
lightest Higgs mass at the tree level.
Hence, this tree level NMSSM contribution to lightest Higgs mass enable
to accommodate easily a SM-like Higgs mass of the range $\sim$ 125 GeV.

As pointed out earlier, the Higgs sector also receives higher order 
perturbative corrections contributed by squark masses and 
trilinear A-terms, and as 
well as by the parameters of the gaugino and Higgsino sectors. 
A very tiny contribution may come from slepton sectors also.  
The total 
extra contribution due to these corrections lead further enhancement of 
Higgs mass. The calculations of these higher order corrections 
exist in the literature 
${\cal O}(\rm{ \alpha_t \alpha_s + \alpha_b \alpha_s})$~\cite{Ellwanger:1993hn,
Elliott:1993bs, Elliott:1993ex,Pandita:1993tg, Pandita:1993hx, 
Ellwanger:2005fh,Degrassi:2009yq, Ender:2011qh, Goodsell:2014pla, 
Muhlleitner:2014vsa} and the NMSSM-specific 
two loop corrections can also be found in
the Ref.\cite{Goodsell:2014pla}.
Including this additional correction to the mass, 
this upper limit of lightest Higgs boson mass can go up to 
$\sim$ 140 GeV with the constraint that 
$\rm{\lambda}$ to be within the perturbative 
limit~\cite{Ellwanger:2006rm,Barbieri:2006bg}.
 
Consequently, the higher order correction to Higgs mass relates the 
other sectors, 
in particular third generation of squarks. Therefore to predict Higgs 
mass, one needs to take into consideration of involved model parameters, 
including very sensitive third generation sector.
On the other hand, perhaps, the well determinations of Higgs sector 
predict about the
range of masses and mixing angles of third generation squarks, namely
top squarks~\cite{Hall:2011aa,Beuria:2015mta}. However, it is to be 
noted here that the different spectrum generators for the NMSSM predict 
different Higgs boson masses separated by few GeV for a given set of 
input parameters and schemes~\cite{Staub:2015aea }.

As described before, in NMSSM the Higgs sector has a non trivial 
dependence on a set of parameters 
Eq.\ref{eq:nmssmparam} ~\cite{Miller:2003ay} which 
eventually affect the NMSSM Higgs phenomenology. Recall that 
generating $\rm{\mu}$ term dynamically leads 
$\rm{\mu_{eff} \sim \lambda v_s}$, which also is connected 
with chargino/neutralino 
masses. Hence, the LEP limit on light chargino mass ($\gsim$ 100GeV)
~\cite{Agashe:2014kda} 
predicts a bound, $\rm{\lambda \ge \mu_{eff}/v_s}$~\cite{Ellwanger:2011sk}. 
Apparently, 
this bound restricts $\rm{\lambda}$ not to be very small unless vev ${v_s}$ 
becomes too large, which is not also phenomenologically favourable   
as the other vevs, $v_u, v_d$ are expected to be also 
around the EW scale. 
{However, there are certain variations of the NMSSM model, for instance,  
the semi-constrained NMSSM where soft masses of scalars, gauginos 
and trilinear A-terms for third generation fermions are assumed 
to be the same at the grand unified scale~\cite{Ellwanger:2014dfa}. 
In this type of scenario, the values of
the $\lambda$ and $\kappa$ can be very small, even much below of 
the ${\cal O}$(0.1) which is completely compatible with the existing 
constraints which will be discussed later.} 
On the other hand,  
running of $\rm{\lambda}$ and $\rm{\kappa}$ from high scale of
Grand Unified Theory(GUT)
to EW scale through renormalization
group equation and 
the requirement of 
perturbative nature of couplings, a bound on them can be 
derived ~\cite{Miller:2003ay} as,
\br
\rm{\lambda^2 + \kappa^2 \lsim 0.6.}
\label{eq:laka}
\er

\section{Scanning of NMSSM parameters }
\label{nugm}
In this section, we try to identify the region that 
offers one of the CP-even Higgs boson as the SM-like ($\rm{H_{SM}}$). 
by carrying out a comprehensive { random } scan 
of NMSSM parameter space.  
We use the package 
NMSSMTool4.3.0~\cite{Ellwanger:2004xm,Ellwanger:2005dv} which
calculates for a given set of input parameters, the masses, the 
couplings and the decay branching ratios of 
all the Higgs bosons along with the spectrum of other SUSY particles.
This code also systematically checks the consistencies 
of the parameters against 
various theoretical as well experimental constraints.  
Experimental constraints include measurements of various low and 
high energy observables predicted within the framework of the NMSSM.

In our scan, we set the following ranges for the relevant
parameters, 
\begin{gather}
\rm{0.1 < \lambda < 0.7;\ \ 0.1<\kappa < 0.7;\ \ 0< A_{\lambda}<2~TeV},
\rm{-2~TeV < A_{\kappa} < 200~GeV}; \nonumber \\   
\rm{1.5 <\tan\beta<30; \ \ 100~GeV<\mu_{ eff}<2~TeV}.
\label{eq:p1}
\end{gather}
{ Here the choice of parameters are phenomenologically motivated
which expected to provide a very different scenario than the MSSM and testable
at the future LHC experiments. The range of parameters are chosen
in such a way that it provides one of the
CP even Higgs boson to be SM-like with a mass around 125 GeV 
and other singlet like Higgs bosons are not too massive.
In our scan very small values of $\lambda$ are
avoided to restrict $\mu_{eff}$ and $\rm v_s$ at the level of 
electroweak scale. 
{ However, as mentioned in sec.2, the values of $\lambda$ and $\kappa$
can be very small even much lower than our considered ranges, in the
context of the certain variation of the NMSSM~\cite{Ellwanger:2014dfa}.
In view of this, our chosen scenario of the paratmeter choices is not 
very general one, represening only a subset of the parameter space.}   
The g-2 constraint favours the  
positive values of $\mu_{eff}$.          
However, this in not a very
general set up of parameters while more wider ranges of parameters 
are also accessible in the context of certain 
variation of NMSSM~\cite{Ellwanger:2014dfa,Agashe:2012zq,Cao:2012yn} 
}.

The soft mass parameters for the left and the right handed squarks from
the first two generations, to which the Higgs boson mass 
are not so sensitive, are set to,
\br
\rm{M_{Q_{1/2}} =M_{U_{1/2}} =M_{D_{1/2/3}}= 1  TeV},  
\label{eq:p2}
\er
while for the third generation squarks, known to have a large 
effect on the Higgs mass, these are varied over the range,
\br
\rm{M_{Q_3}=M_{U_3}=300-3000~GeV}.
\label{eq:p3}
\er
to consider a wide range of values from them. However, as we know, 
because of radiative contributions to the mass of the Higgs boson, the 
masses of the top squarks 
and the corresponding mixing are important and hence A-term plays
an important role. With a goal to achieve a CP even Higgs state close
to the mass of the SM Higgs, we vary $|\rm A_t|$ from small to large values 
as,
\br
\rm{A_t = -4 ~{\rm TeV} ~{\rm to}~ +4 ~{\rm TeV}},
\label{eq:p4}
\er
and setting other third generation trilinear A-terms as, 
\br
{\rm 
 A_b=2~{\rm TeV}}; \ \ \rm{and} \ \  
\rm{A_{E_3}=1~TeV},
\label{eq:p5}
\er
while 
keeping all other remaining trilinear A terms to zero. 
While scanning parameters,
we test all constraints including theoretical ones pertaining to  
vacuum stability. Moreover, the precision measurements at the LEP experiments 
restrict the new physics models through the measurement 
of $\rm{M_W}$, $\rm{\sin^2\theta_{eff}}$,
the $\rm{\rho}$ parameter and also via the invisible decay width of the 
Z boson.
In addition, the constraints due to various low energy observables from 
flavor physics such as
$\rm{B \to X_s \gamma}$, $\rm{B_s \to \mu^+\mu^-}$ 
and $\rm{B \to \tau^+ \nu_\tau}$
and also the mass differences $\rm{\Delta M_d}$ 
and $\rm{\Delta M_s}$ are checked
in the process of scanning of the parameters.
Various existing bounds on sparticle masses and production
cross sections that are obtained from LEP, the Tevatron and from the LHC
experiments are also imposed. The details of all these constraints
can be found in \cite{Ellwanger:2006rn}.
Notably, among various experimental 
constraints, the most crucial ones are from 
the Planck \cite{Ade:2013zuv}  data and the muon anomalous
magnetic moment \cite{Bennett:2004pv}. In this model the lightest 
neutralino($\tilde\chi_1^0$) is assumed to be 
the lightest-supersymmetric particle (LSP)   
and is also a DM candidate. NMSSMTools interfaced with micrOMEGAs
~\cite{Belanger:2013oya,Belanger:2005kh},  
calculates the relic density.  
Recent data from the PLANCK experiment  
\cite{Ade:2013zuv} concludes the relic density to be,
\br
\rm{\Omega^2 h = 0.1187 \pm 0.0017}.
\label{eq:dmc}
\er
In the NMSSM, the DM solution is tested with the
measured data as shown above at the 10\% level. The most natural solution
to the DM problem is via the bulk annihilation channel which requires
lighter electroweak gauginos and sleptons. The impact of DM 
constraints on the NMSSM are discussed by 
many authors \cite{Vasquez:2012hn,Ellwanger:2014dfa,Belanger:2005kh,AlbornozVasquez:2012px,Kozaczuk:2013spa,LopezFogliani:2012yq,Cao:2011re,Das:2010ww,Hugonie:2007vd,LopezFogliani:2007sc,Cerdeno:2007sn,Gunion:2005rw,Enberg:2015qwa}.
The parameters in the gaugino sector, $\rm {M_1}$ and $\rm{ M_2}$, 
the U(1) and the SU(2) gaugino masses, respectively are expected not to have 
any large effects on Higgs masses, but are important in calculating
the dark mater relic density, Eq.~\ref{eq:dmc}.
Therefore, we vary the values of gaugino masses 
$\rm{M_1, M_2}$ while  keeping $\rm{M_3}$, which is 
{ very close to the value of the} gluino mass to a fixed value,
\br
\rm{50~GeV<M_1 < 1~TeV}; \ \  \rm{50~GeV< M_2 < 1~TeV}; \ \  \rm{M_3=1.2~TeV}.
\label{eq:p6}
\er
The precisely 
measured value of the muon anomalous magnetic moment, 
\br
\rm{a_\mu = \frac{(g_\mu -2 )}{2} }
\er
is considered to be a very strong constraint for new physics. 
The total SM contribution to $\rm{a_\mu}$ is estimated with an uncertainty 
predominantly due to
the hadronic contributions \cite{Jegerlehner:2007xe}. The discrepancy 
between the measured value and the one  
predicted by the SM is found to be~\cite{Bennett:2004pv}, 
\br
\rm{\delta a_\mu = (28.7 \pm 8.0)\times 10^{-10}}.
\label{eq:gm2}
\er 
This implies the measured value is 3$\sigma$ away from the SM predicted 
value. In SUSY model, there are additional contributions due to the 
presence of sparticles in loops 
which provide an explanation for this excess~\cite{Endo:2013bba,Moroi:1995yh,Gnendiger:2013pva,Fargnoli:2013zia}. In the NMSSM, the effect of $g_\mu$-2
constraint on its model parameter space has been discussed 
in \cite{Domingo:2008bb}.
As we know, the main SUSY contribution to
$\rm{a_\mu}$ comes at the one loop level involving smuon-neutralino
and sneutrino(muon)-chargino diagrams. However, this SUSY 
contribution at the one loop
level is determined by the sign of the $\rm{\mu}$-term and the 
value of $\rm{\tan\beta}$. In order
to have a reasonable contribution to $\rm{\delta a_\mu}$ from 
these additional one loop SUSY diagrams, the masses of the sparticles, 
in particular of the smuons  
are favoured to be as light as $\sim$~100 GeV and 
$\rm{\tan\beta \sim} $ 10-15~\cite{Domingo:2008bb}. The slepton masses 
are not directly 
coupled with the Higgs sector,
but plays a crucial role in determining the anomalous magnetic moment
of the muon as discussed before. In order to have a parameter space 
consistent with $g_\mu$-2 constraint
we assume low values of the first two generations of
sleptons~\cite{Domingo:2008bb} i.e,
\br
\rm{M_{L_{1,2}}=100~GeV; \ \ M_{E_{1,2}}=100~GeV}.
\label{eq:p5}
\er
The Higgs contribution to $\rm{\delta a_\mu}$  comes via 2-loop 
diagrams\cite{Krawczyk:2002df,Heinemeyer:2004yq} and is found to be
negligible both in the SM and in the MSSM
because of the { higher values of the Higgs masses.} 
On contrary, in the NMSSM rather light Higgs states decoupled from 
fermions and bosons are viable, even with a few GeV mass. The one loop 
diagrams involving these lighter Higgs states are found to be 
potential sources contributing 
to $\rm{\delta a_\mu}$~\cite{Gunion:2005rw,Domingo:2008bb}.
For instance, for lower values of the Higgs mass $\lsim$10 GeV, for 
both CP even and CP odd
states, contributions appear to be substantial 
within 2$\rm{\sigma}$ of the central value of $\rm{\delta a_\mu}$
given in Eq. \ref{eq:gm2} 
for large 
values of $\rm{\tan\beta}$. In our scan, 
we take into account the impact of $g_\mu$-2 constraint which has 
some effects on 
the parameter space as will be discussed later.
              
In the NMSSM, out of three CP even Higgs bosons, one is required to 
be SM-like. In addition, the couplings of the SM-like
Higgs boson to fermions and the gauge bosons should be consistent with the 
current measurements by 
both the CMS and the ATLAS experiments \cite{Khachatryan:2014jba}.
As shown in Eq.\ref{eq:hcoup}, \ref{eq:acoup} and \ref{eq:sumrule1},
the couplings of the Higgs boson with fermions and the gauge boson are 
basically a scaling of the corresponding SM couplings by a factor 
called the reduced couplings which are essentially the ratios of 
$\rm{ { g_{H_i ff}}/{g_{H_i ff}^{SM}} (=C_f)}$ or 
$\rm{{g_{H_i VV}}/{g_{H_i VV}^{SM}} (=C_V)}$. 
Similarly, one can also have reduced effective couplings of Higgs with 
gluons and photon, say $\rm{C_g}$ and $\rm{C_\gamma}$, respectively. 
At the LHC, both the ATLAS and the CMS collaborations examined the 
the Higgs couplings 
following the strategy discussed in Ref.\cite{Heinemeyer:2013tqa}.  
By performing a very detailed analysis, the allowed ranges of the 
reduced couplings 
are reported at 95\% C.L. In Table \ref{tab:coupling} allowed 
ranges of these reduced couplings measured by the CMS 
experiment~\cite{Khachatryan:2014jba} are presented.
{  Note that the CMS and ATLAS experiments do not measure the
Higgs couplings directly. However, by measuring the signal strength, 
the factors by which the SM couplings are scaled can be constrained.
Here these scale factors are defined to be the reduced couplings which are
constrained by both the LHC experiments.}
It is to be noted that the SM Higgs production cross sections 
in various related modes 
are used
in deriving these limits. 
In our study, while scanning the parameter space,  we impose these 
constraints on the reduced couplings for the SM-like CP even Higgs boson. 
Similarly, from negative searches of Higgs at the LEP experiment,
the reduced couplings of the SM-like Higgs boson with W/Z ($\rm{ C_v=\xi}$)
also get constrained \cite{Schael:2006cr} and our 
scan has been subjected to 
these constraints as well.
Furthermore, in the NMSSM, possibly the SM-like Higgs boson can decay via  
non-standard modes which will be discussed in later section.
\begin{table}
\begin{center}
\begin{tabular}{|l|ll|}
\hline
 & Lower & Upper \\
\hline
$C_t$ & 0.97 & 2.28 \\
$C_b$ & 0.00 & 1.23 \\
$C_V$ & 0.66 & 1.23 \\
$C_g$ & 0.52 & 1.07 \\
$C_\gamma$ & 0.67 & 1.33 \\
\hline
\end{tabular}
\caption{The allowed ranges of reduced couplings of the
SM-like Higgs with fermions and the gauge bosons and 
also the effective couplings with the gluons and
the photons at 95\%C.L.
These are obtained by CMS experiments~\cite{Khachatryan:2014jba}.}
\label{tab:coupling}
\end{center}
\end{table}
The total BR in these non SM modes, namely $\rm BR_{BSM}$ of the SM 
like Higgs boson is also restricted from above from the current Higgs data. 
The CMS measurements set the upper limit of non-SM branching ratio 
of the SM-like Higgs to be \cite{Khachatryan:2014jba},
\br
\rm{ BR_{BSM} \le 0.57 ~ at~ 95\%. C.L}.
\label{eq:brns}
\er
This limit is incorporated in our parameter scan.

Eventually, after playing with the NMSSM model parameters, 
we identify the region of parameter space
allowed by all theoretical and experimental
constraints as described.above, 
which yields one of the relatives light  
CP even Higgs bosons, either $\rm H_1$ or $\rm H_2$, to be 
SM-like with its mass in the range,
\br
\rm {\mh/\mhh \sim m_{\hsm} = 125.02 \pm 3~{\rm GeV}}.
\label{eq:hsm}
\er 
It is always observed that the heaviest CP even neutral Higgs, $\rm H_3$
tends to be massive $\sim$ 200 GeV or more.
Thus, we focus our attention on two distinct regions 
presenting the following two scenarios: 
\\
$\bullet$ Case A: 
the second lightest CP even Higgs, $\rm H_2 \sim \rm{\hsm}.$ 
\\
$\bullet$ Case B: the lightest CP even state, $\rm H_1 \sim \rm{\hsm}$.
\\
In the subsequent sections we discuss various features of these two cases
and the relevant phenomenology in the context of 
ongoing LHC experiment with a center of mass energy of 13 TeV.
\subsection{Case A: $\rm H_2 \sim \rm{\hsm}  $ }
\label{H2SM}
In this section we discuss the salient features of the NMSSM Higgs sector 
corresponding to the parameter space 
which leads to the second lightest CP
even Higgs boson, $\rm{H_2}$, as SM-like.   
\begin{figure}[ht]
\centering
  \includegraphics[width=0.60\textwidth]{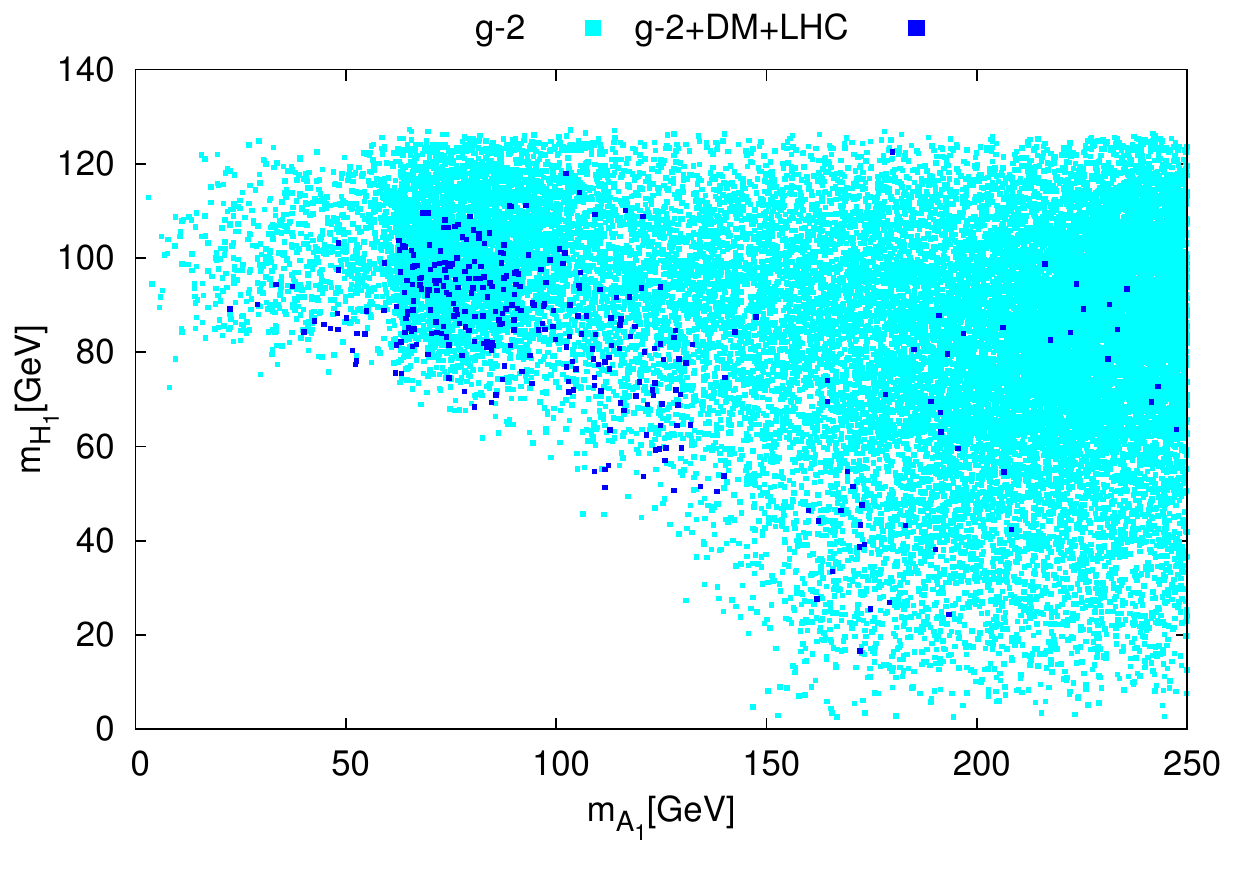}
\caption{Range of $\rm \ma$ and $\rm \mh$ for case(A), allowed  
by all theoretical and experimental constraints 
as discussed in Sec.3 plus $g_\mu-2$ constraint(cyan) and 
then adding the LHC and DM constraints(blue).}  
\label{fig:mh1ma1}
\end{figure}
Here, the lightest CP even 
Higgs($\rm{H_1}$) lighter than $\rm{ H_2}$ by definition. In addition 
the lightest CP odd Higgs boson($\rm A_1$) is also lighter 
than $\rm H_2$ in a certain region of parameter space. Both  
$H_1$ and $A_1$ are dominantly singlet like.   
In Fig.~\ref{fig:mh1ma1}, we present the possible ranges of $\rm \mh$ and
$\rm  \ma$ when $H_2$ is SM-like~\cite{King:2014xwa}. 
In this figure the allowed points (cyan) are 
result from imposing only the $g_\mu$-2 limit
(Eq.\ref{eq:gm2})  
along with all other theoretical and experimental constraints 
as described in the previous section. Subsequently, adding constraints 
from the DM relic density 
(Eq.~\ref{eq:dmc}) and from the measurements of the Higgs couplings at the LHC 
(Table~\ref{tab:coupling}), the more restricted 
region(blue) is obtained. It is interesting to note that, even after  
including all such constraints, $H_1$ or $A_1$ 
may turn out to be very light, $~\sim$ 20 GeV, although 
there are only few points~\cite{Dermisek:2008uu,Louis:2015lma}.
Moreover, this figure clearly indicates that for a good fraction 
of the parameter space, both $\rm \mh$ and $\rm \ma$  
are found to be less than the half of the mass of $\rm H_2$,
thus allowing the latter decay in the following ways,
\br
\rm{H_2} &\to& \rm{H_1 H_1 }, \nn \\
     &\to& \rm{ A_1A_1 }.
\label{eq:h2dk}
\er
It is to be noted that these lighter singlet states $\rm A_1$ and 
$\rm H_1$ are not yet ruled out   
by any collider experiments.  
These singlet like states escape detection in collider experiments  
because of their suppressed couplings to gauge bosons and fermions.
\begin{figure}[ht]
 \centering
  \includegraphics[width=0.45\textwidth]{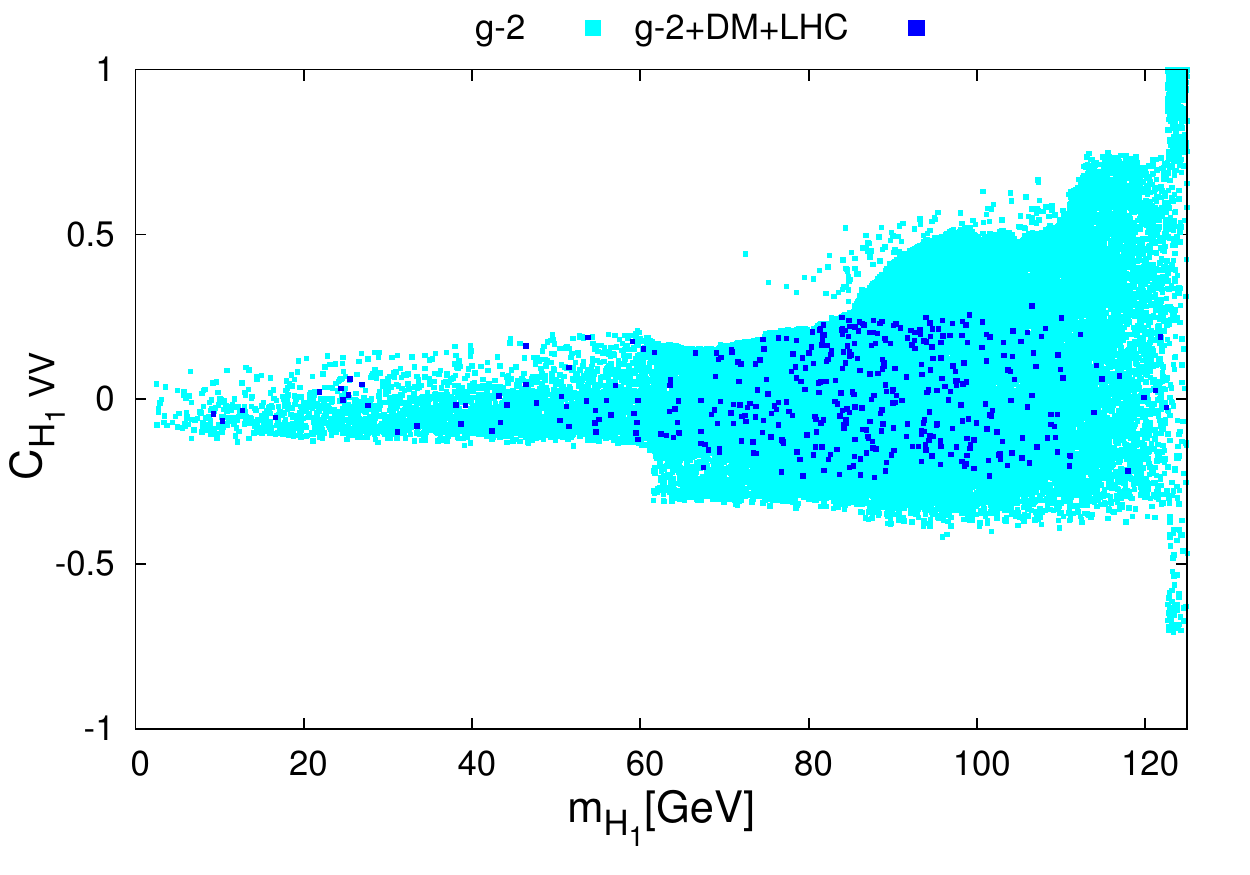}
  \includegraphics[width=0.45\textwidth]{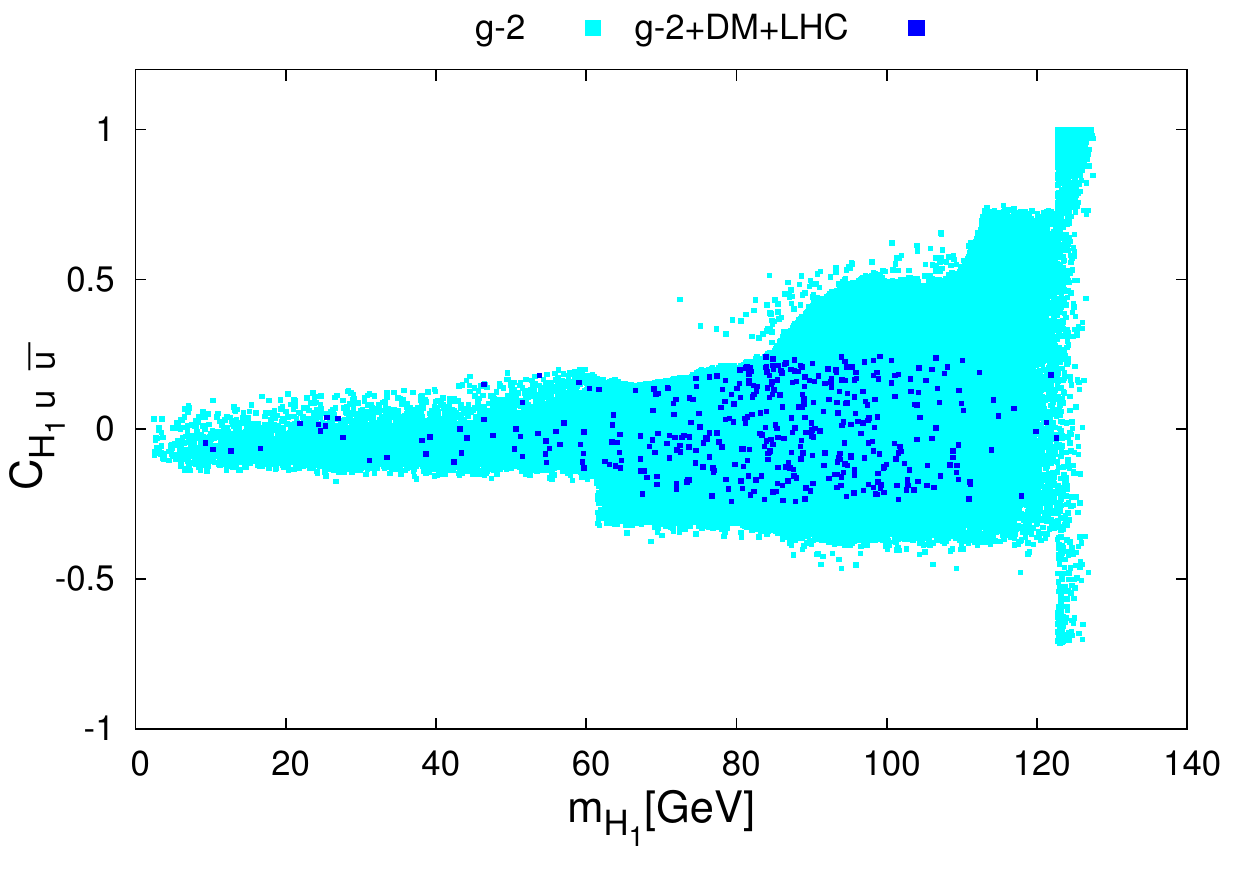}
   \includegraphics[width=0.45\textwidth]{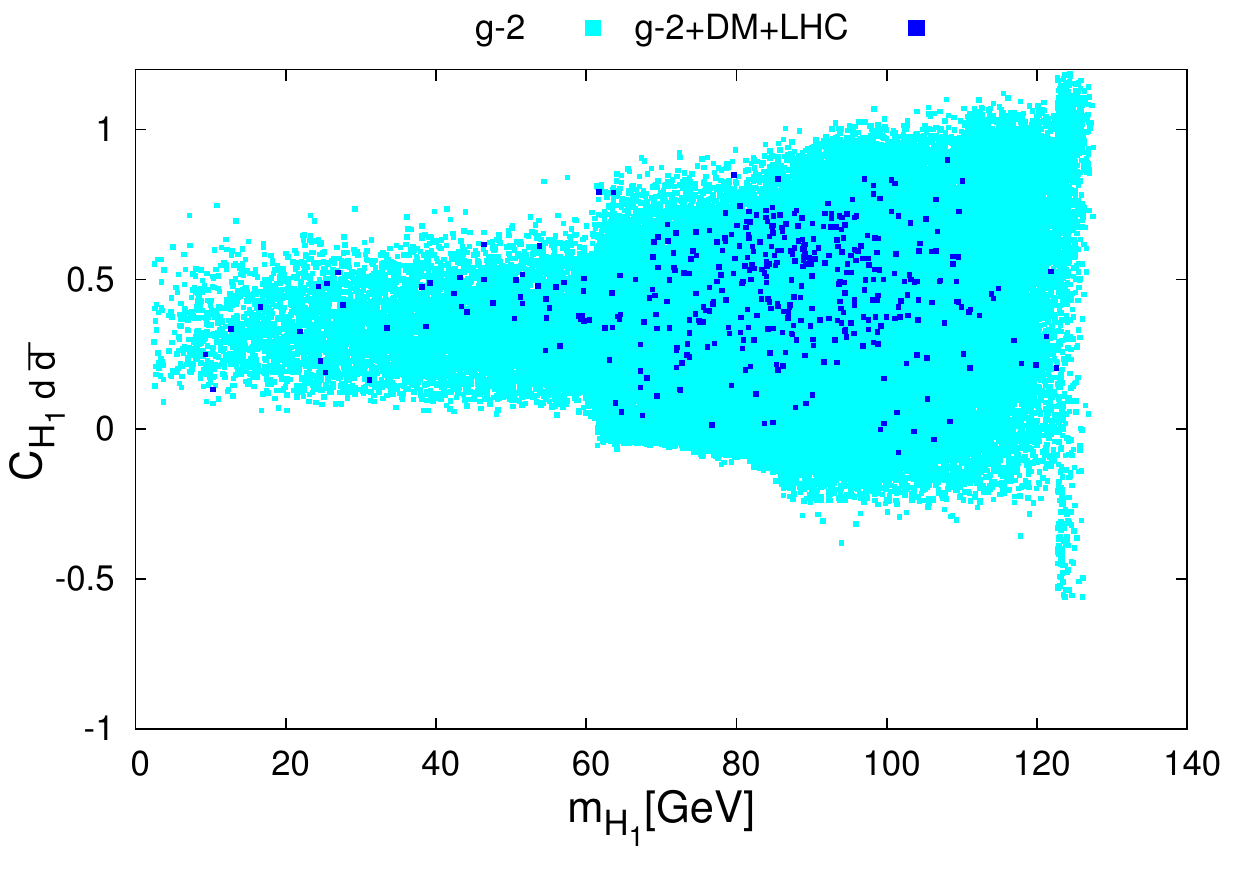}
\caption{Reduced couplings of $\rm H_1$ with gauge bosons(V=W/Z) 
and fermions (up and down types) for case(A). 
Values are allowed  by mentioned constraints
( conventions are the same as in Fig.~\ref{fig:mh1ma1}.)}
  \label{fig:h1coupl}
\end{figure}
It can be realized by looking at Figs.\ref{fig:h1coupl} and 
~\ref{fig:a1couplh2} which present 
for a wide range of $\mh$, 
the reduced couplings of $\rm H_1$ and $\rm A_1$, respectively,
with gauge bosons, up and down-type 
quarks. 
At lower masses for $\rm H_1$, allowed values of these reduced 
couplings initially appear 
to be very small and then 
go up with increasing $\rm{m_{H_1}}$. However, 
LHC constraints, in particular
restrictions on reduced couplings of the SM-like Higgs boson  
as shown in Table~\ref{tab:coupling}, favor smaller values, except 
for the case $\rm{C_{H_1d\bar d}}$
which receives some enhancement for higher values of
$\rm{ \tan\beta}$ (see Eq.~\ref{eq:hcoup}). 
In case of $\rm{A_1}$, the pattern of variation of reduced 
the couplings with $\rm{m_{A_1}}$ is a  
little different since $\rm A_1$ physical state contains a finite 
fraction of doublet 
component and the reduced couplings increase for higher values of 
$\tan\beta$.
As observed in Fig.~\ref{fig:a1couplh2}.
the reduced coupling of $\rm{A_1}$ to d-type quark is somewhat  
larger than its coupling to u-type quark.
Consequently, this behaviour of the reduced couplings 
significantly affects the Higgs phenomenology in the colliders.
These are discussed in the next section. 
\begin{figure}[H]
 \centering
  \includegraphics[width=0.45\textwidth]{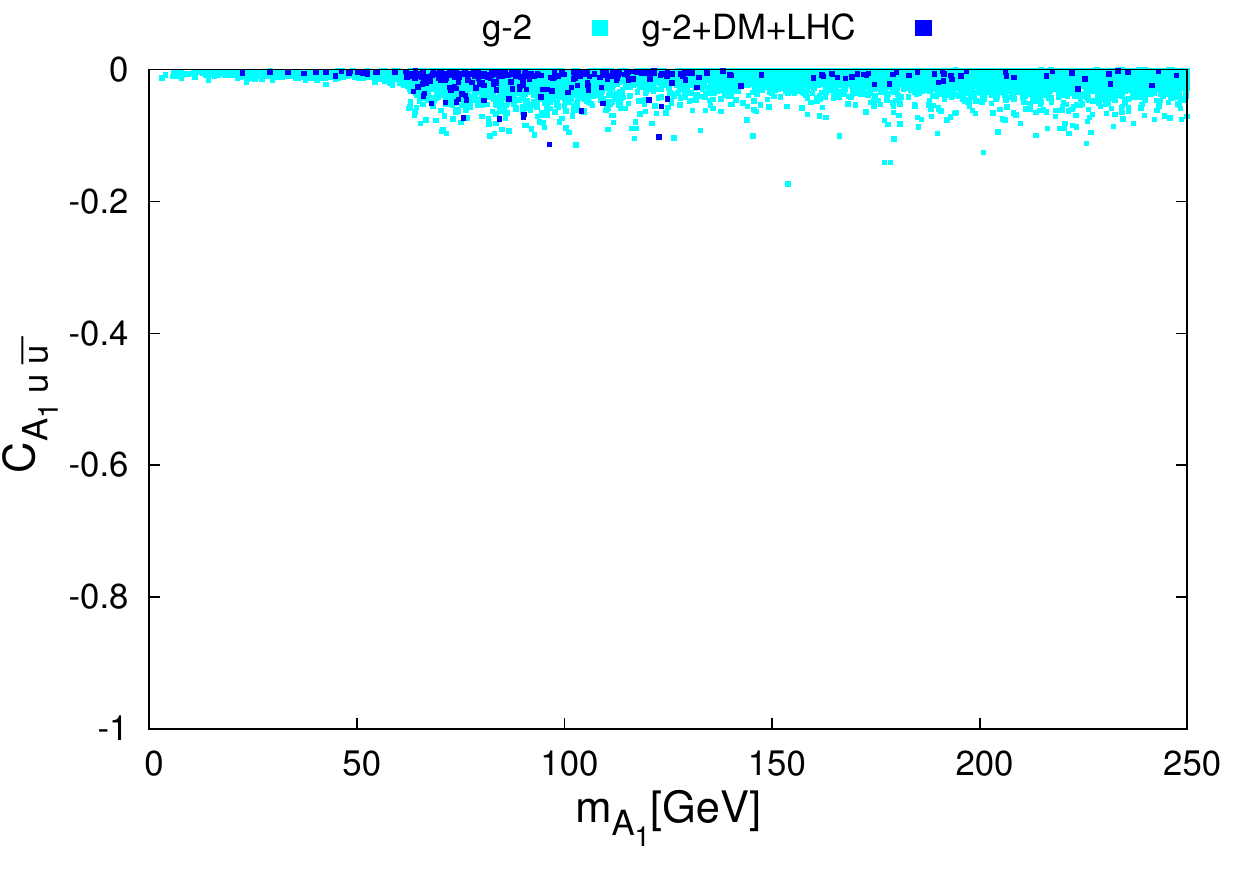}
  \includegraphics[width=0.45\textwidth]{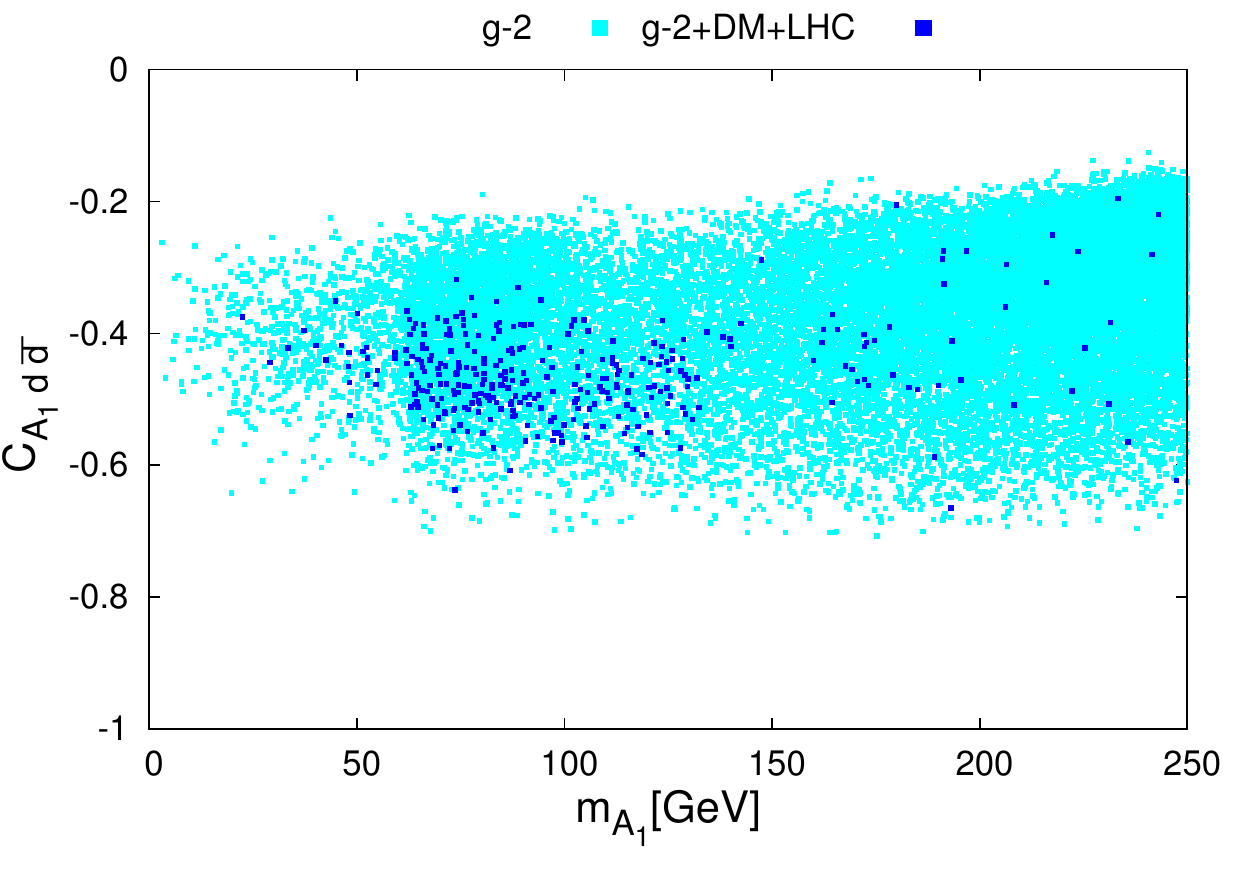}
\caption{Reduced couplings of $\rm A_1$ with quarks 
(up and down types). Other conventions are same as in 
Fig.~\ref{fig:h1coupl}.}
\label{fig:a1couplh2}
\end{figure}

\subsection{Case B: $\rm{H_1} \sim \rm{\hsm} $ }
\label{H2SM}
Following a similar strategy as described above, we obtain a substantial 
region 
of parameter space with the lightest CP even state $\rm{H_1}$ as 
the SM-like Higgs boson.
\begin{figure}[ht]
 \centering
  \includegraphics[width=0.52\textwidth]{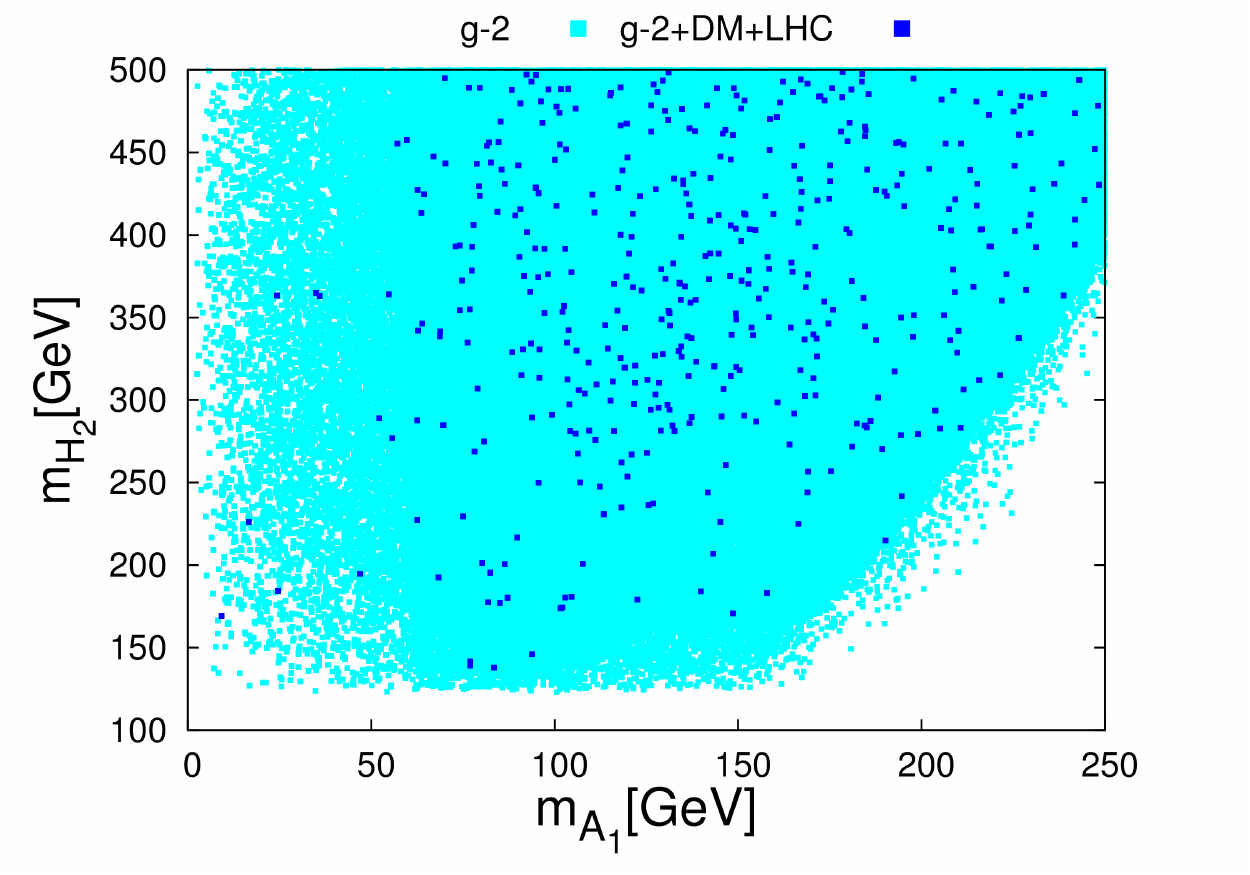}
  \caption{Allowed range of $\ma$ and $\mhh$ for case (B). Constraints
are the same as in Fig.~\ref{fig:mh1ma1}.}
  \label{fig:ma1mh2}
\end{figure}
Obviously, in this scenario, $\rm{m_{H_2}} \gsim \rm{m_{H_1}\sim m_{H_{SM}}}$ 
and the mass of $\rm{A_1}$ can be lighter or heavier 
than $\rm{H_1}$. In fact, a 
significant region of parameter space exists even 
after imposing all constraints where it is observed that $\rm A_1$ is 
light, around $\sim 50-100$ GeV and it can 
turn out to be even  
lighter $\sim 10$~GeV for few points( see   
Fig.~\ref{fig:ma1mh2}), where allowed 
ranges of $\ma$ and $\mhh$ are presented for this scenario. 
This figure indicates 
that the available  
range of $\ma$ can be lighter than $\mh/2$, whereas $\mhh$
may extend to very large ($\sim 500$~GeV) values. 
Hence the following decay channels, 
\br
\rm{H_2} &\to& \rm{H_1 H_1, A_1 A_1}, \nonumber \\  
\rm{H_1} &\to& \rm{A_1 A_1}, 
\label{eq:h12dk}
\er
open up with a reasonable branching ratio depending on the region of the 
parameter space. 
These decay channels may potentially provide additional indirect sources of 
productions of $\rm{H_1}$ as well as of $\rm{A_1}$ at the LHC mediated
by triple Higgs bosons couplings.
\begin{figure}[H]
 \centering
  \includegraphics[width=0.45\textwidth]{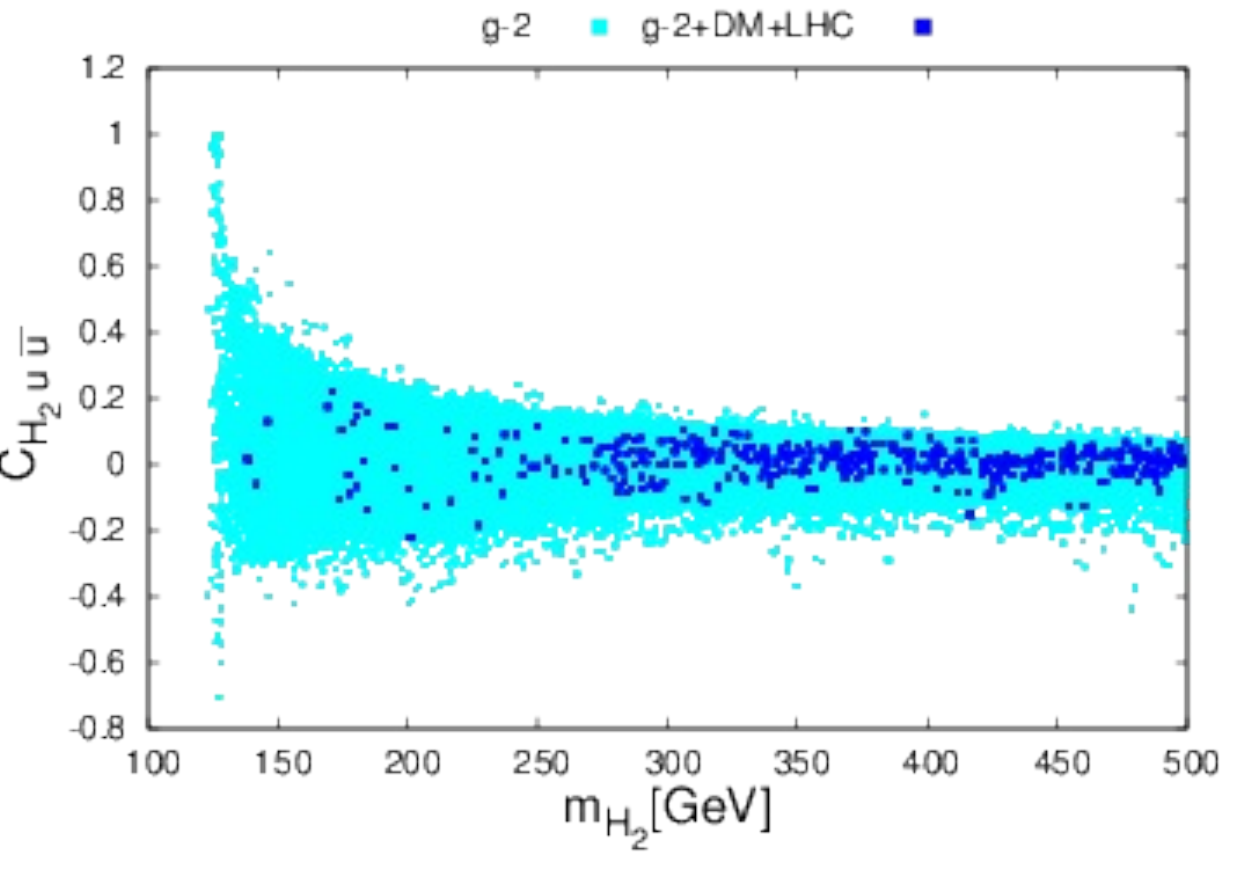}
  \includegraphics[width=0.45\textwidth]{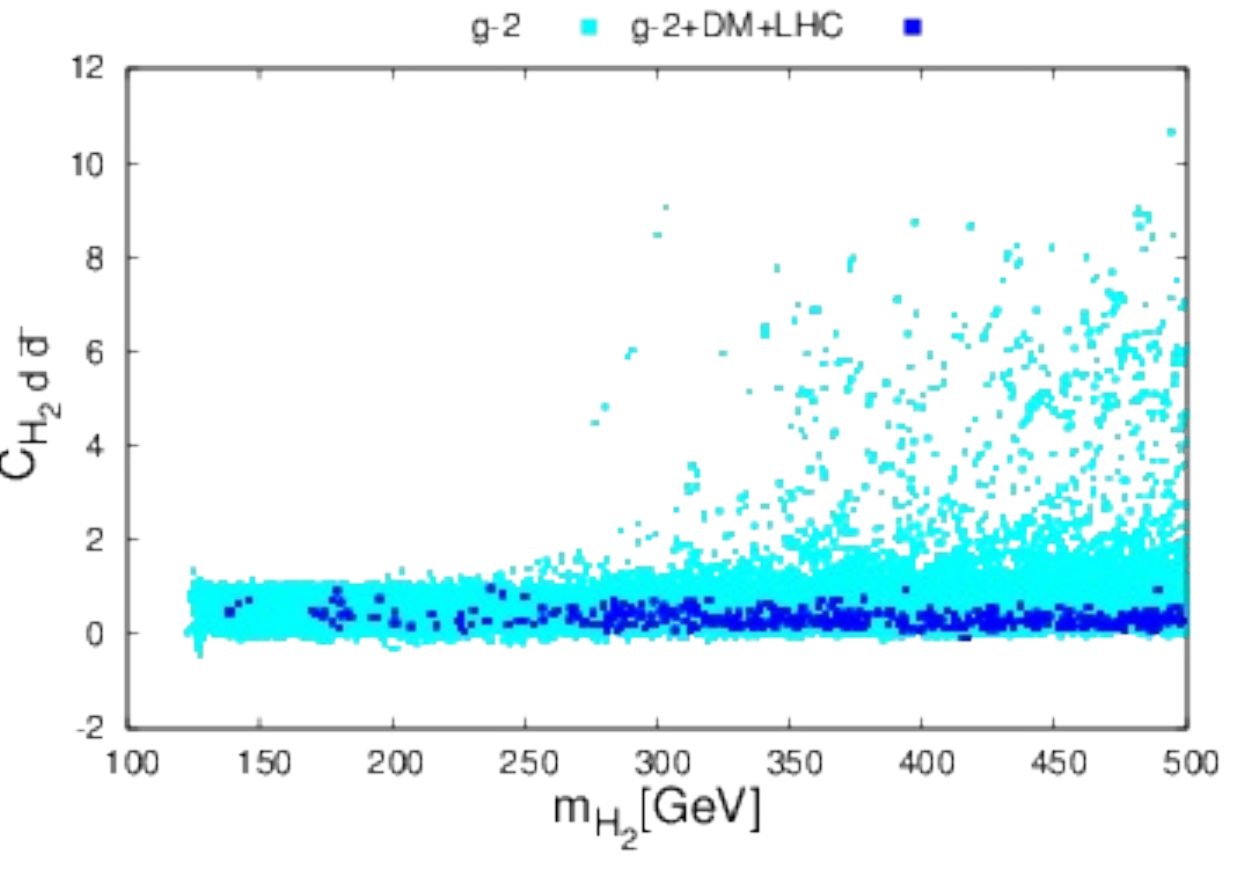}
   \includegraphics[width=0.45\textwidth]{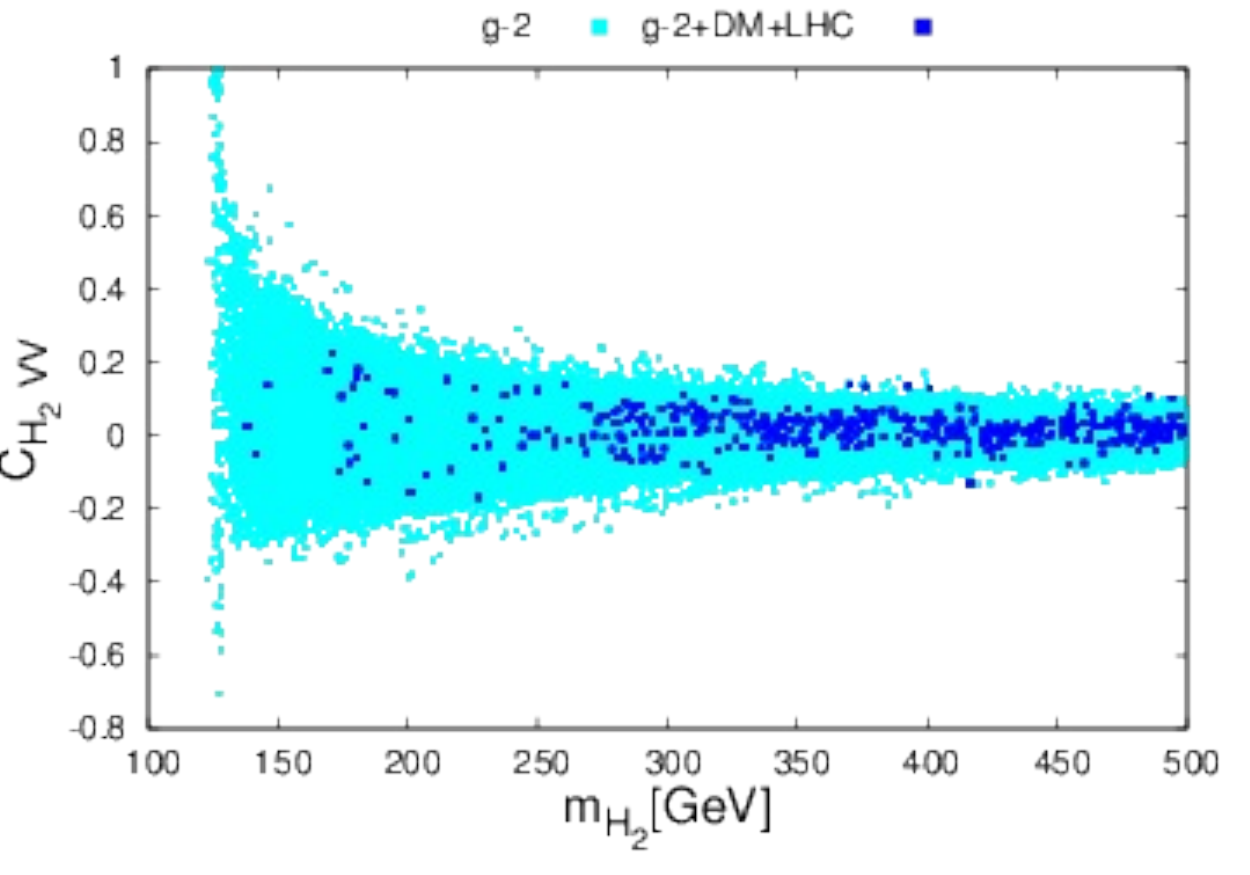}
  \caption{Reduced couplings of $\rm H_2$ with gauge bosons (V=W/Z) and
quarks( up and down types) for case(B). Constraints are the same as 
Fig.~\ref{fig:mh1ma1}.}
  \label{fig:h2coup}
\end{figure}
\begin{figure}[ht]
 \centering
  \includegraphics[width=0.45\textwidth]{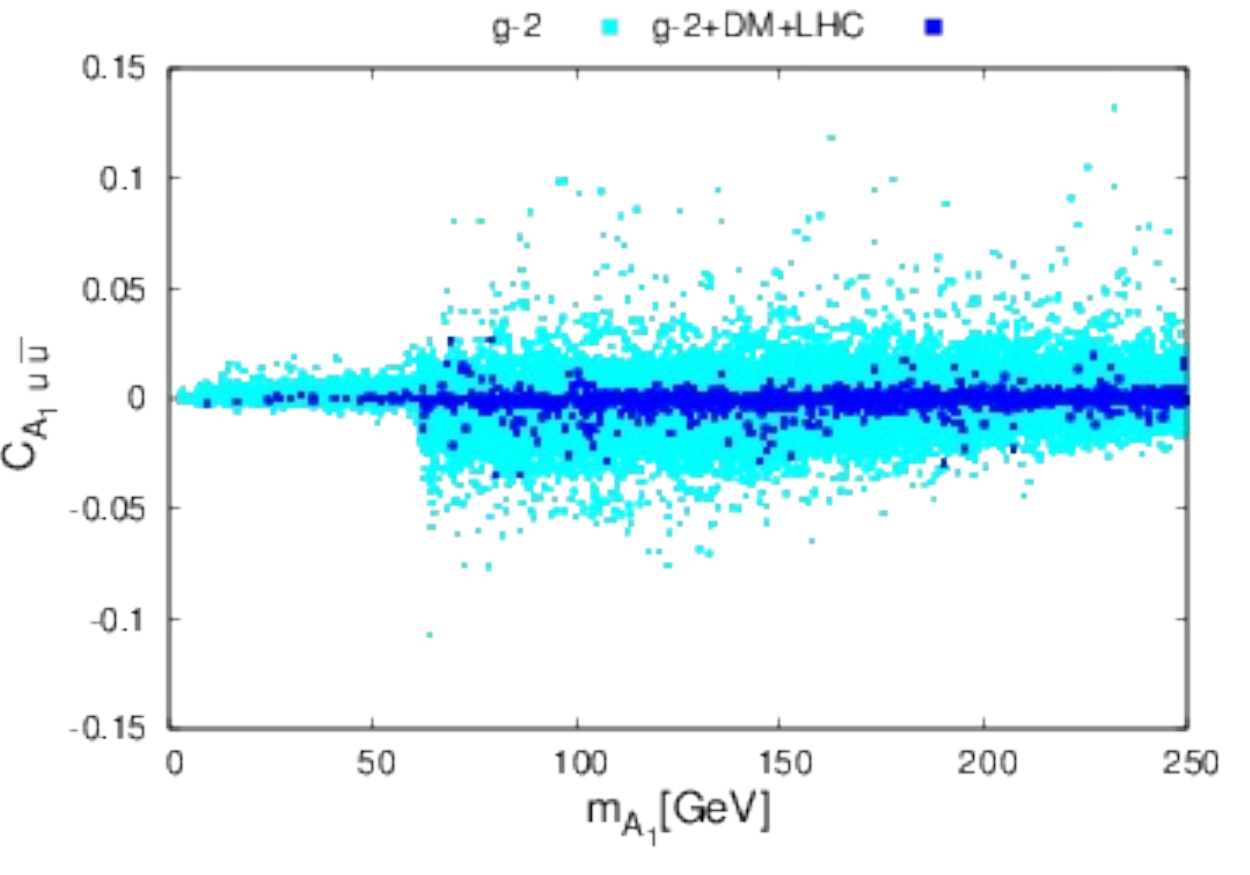}
  \includegraphics[width=0.45\textwidth]{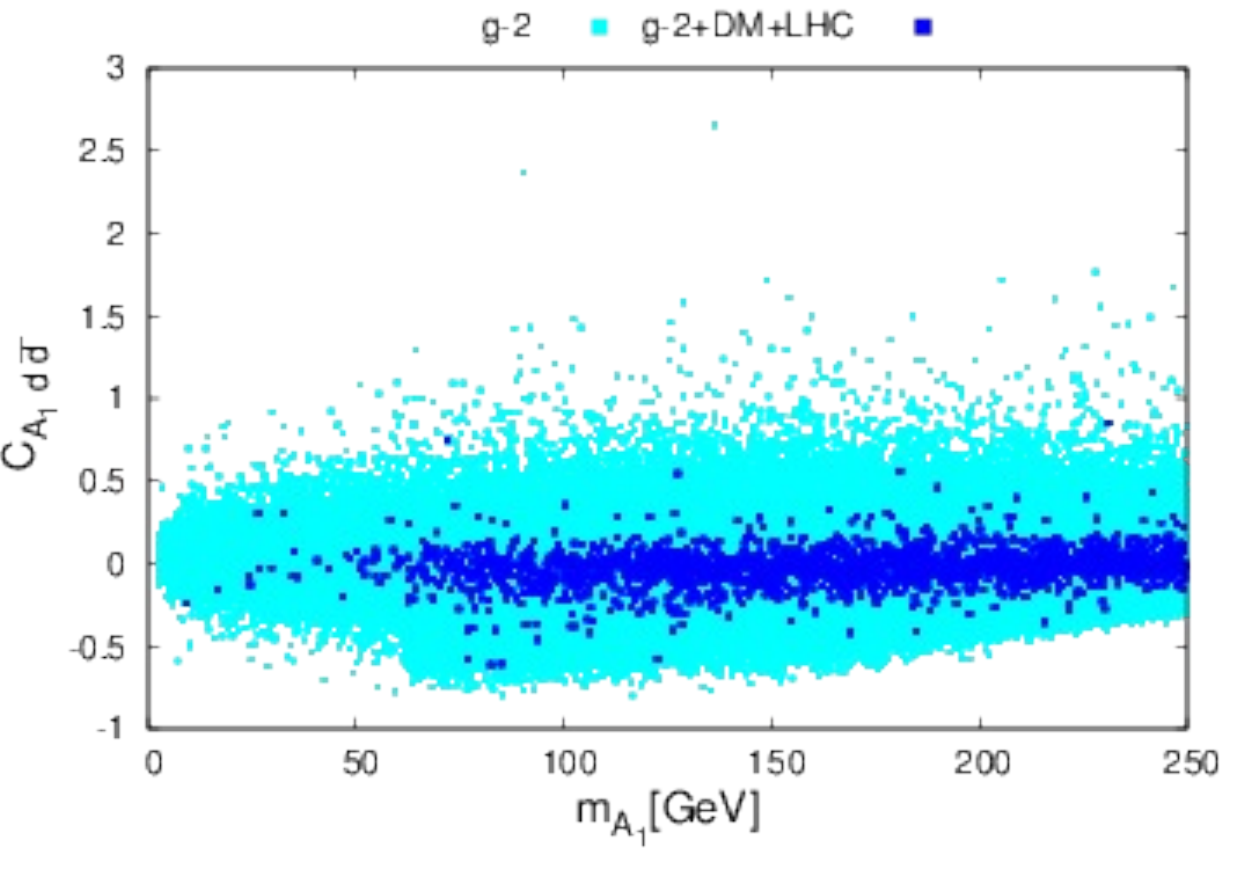}
  \caption{Same as in Fig.~\ref{fig:a1couplh2}, but for case(B).}
  \label{fig:a1coup}
\end{figure}
As before, in order to understand the phenomenology in this scenario 
of non SM singlet-like $\rm{H_2}$ and $\rm{A_1}$, we investigate 
the coupling strengths of 
these Higgs states. 
Since $\rm{H_1}$ is SM-like, and it predominantly couples to gauge bosons, 
obviously the sum rule (Eq.~\ref{eq:sumrule2}) predicts suppression of  
the corresponding reduced couplings of $\rm{H_2}$, as shown in 
Fig.~\ref{fig:h2coup} (bottom panel). 
Notice that the reduced couplings of $\rm{H_2}$ with d-type quark  
are reasonably large, $\sim$2-10 depending on the masses,
but additional DM and LHC constraints restrict those to lower values.  
The variation of the reduced couplings of $\rm A_1$ 
is also studied and is shown in Fig. \ref{fig:a1coup}. 
The reduced 
couplings of $\rm{A_1}$ to the u-type quark is almost negligible whereas  
with the d-type quark it has finite values. {Moreover, 
because of the absence of $\rm A_1$ coupling to gauge bosons,    
direct production of $\rm A_1$ in lepton colliders and via vector boson fusion 
in hadron colliders are suppressed. Hence, the low masses of $A_1$ are not 
excluded by LEP experiment. However, in hadron colliders $A_1$ can be 
produced via gluon gluon fusion or b -$\rm \bar b$ annihilation and 
due to the suppressed couplings of 
$A_1$ with u-type quarks, the latter process is the comparatively 
dominant one.}  
 
Armed with this knowledge about the spectrum of allowed masses and 
the couplings of the light neutral non SM Higgs bosons for 
the two scenarios, (A) and (B), we now discuss their decay patterns 
and followed by their phenomenological implications at the LHC.
\subsection{Higgs decays: $\rm{ H_1, A_1, H_2}$}
\label{lhcsearch}
\begin{figure}[H]
 \centering
  \includegraphics[width=0.45\textwidth]{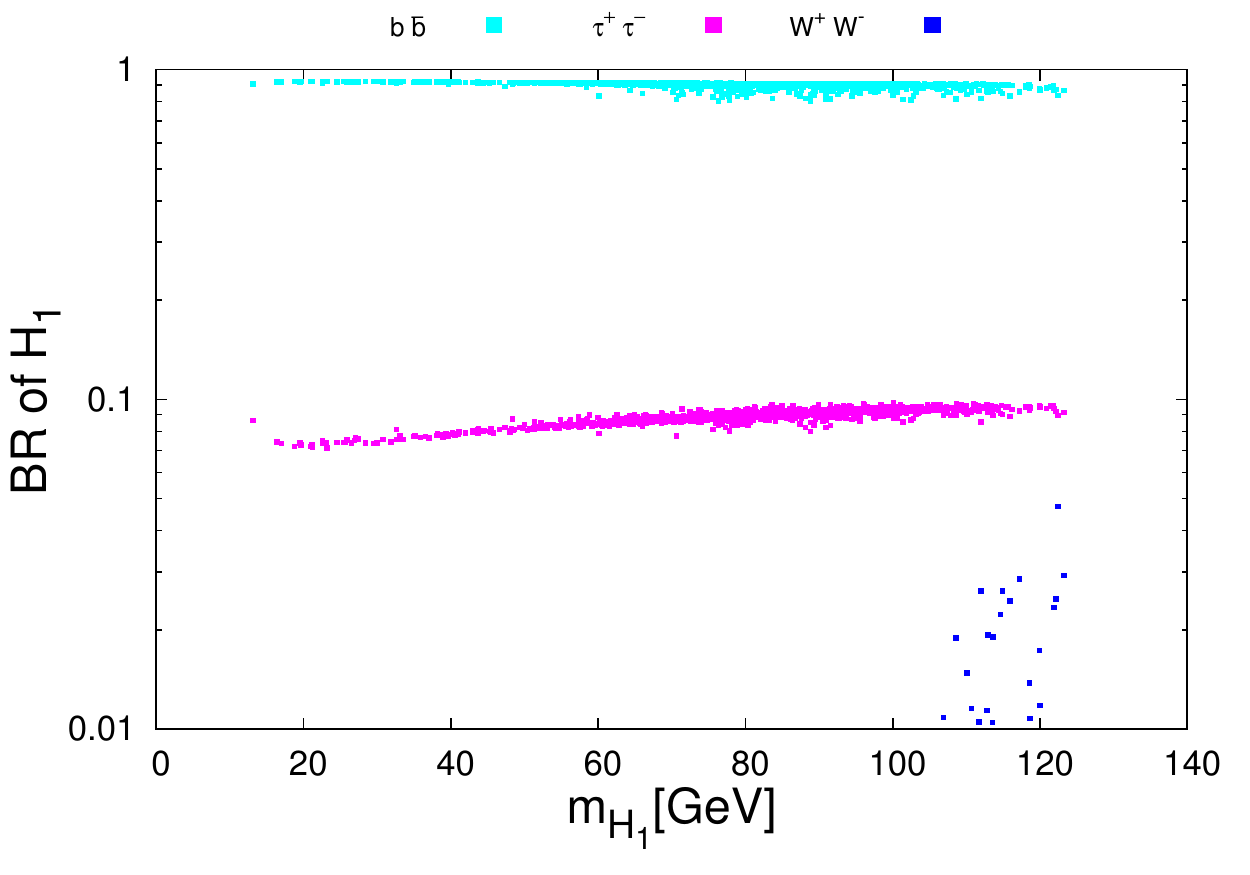}
 \includegraphics[width=0.45\textwidth]{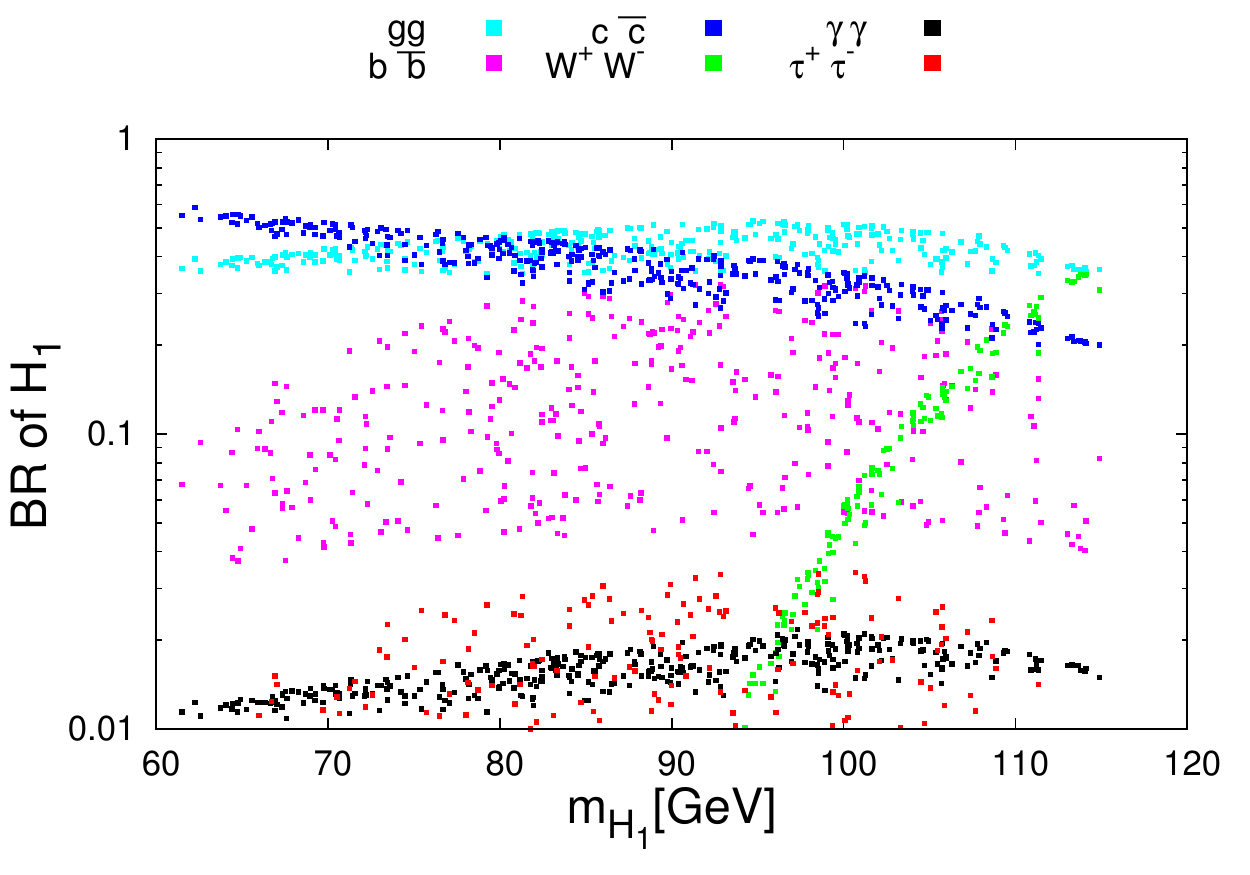}
\caption{BR of $\rm H_1$ in different channels for case(A) and 
allowed by all constraints including $g_\mu$-2, LHC and DM.}
  \label{fig:brh1dkh2}
\end{figure}

\begin{figure}[H]
 \centering
 \includegraphics[width=0.45\textwidth]{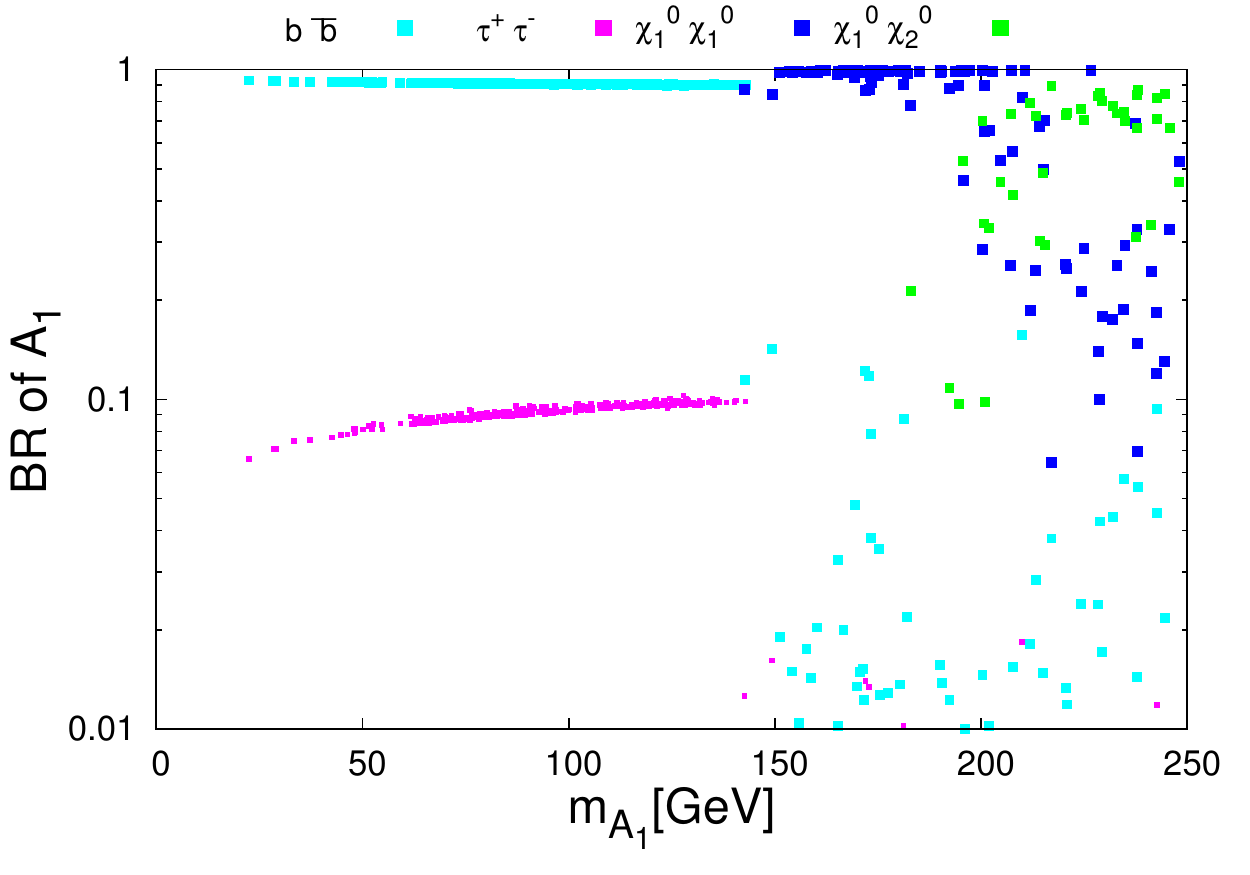}
\includegraphics[width=0.45\textwidth]{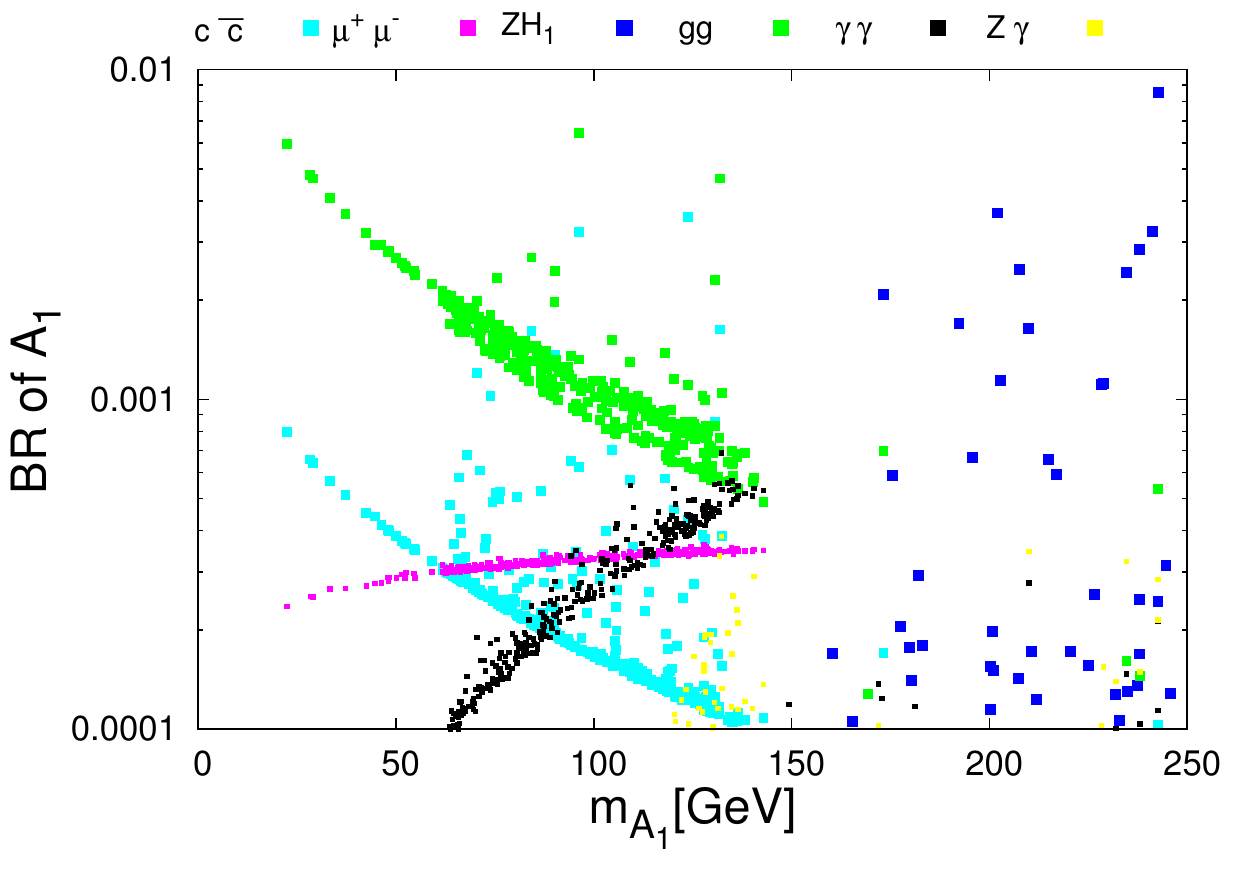}
\caption{BR of $A_1$ for case(A), same as in Fig.~\ref{fig:brh1dkh2}.}
  \label{fig:bra1dkh2}
\end{figure}

\begin{figure}[H]
 \centering
  \includegraphics[width=0.45\textwidth]{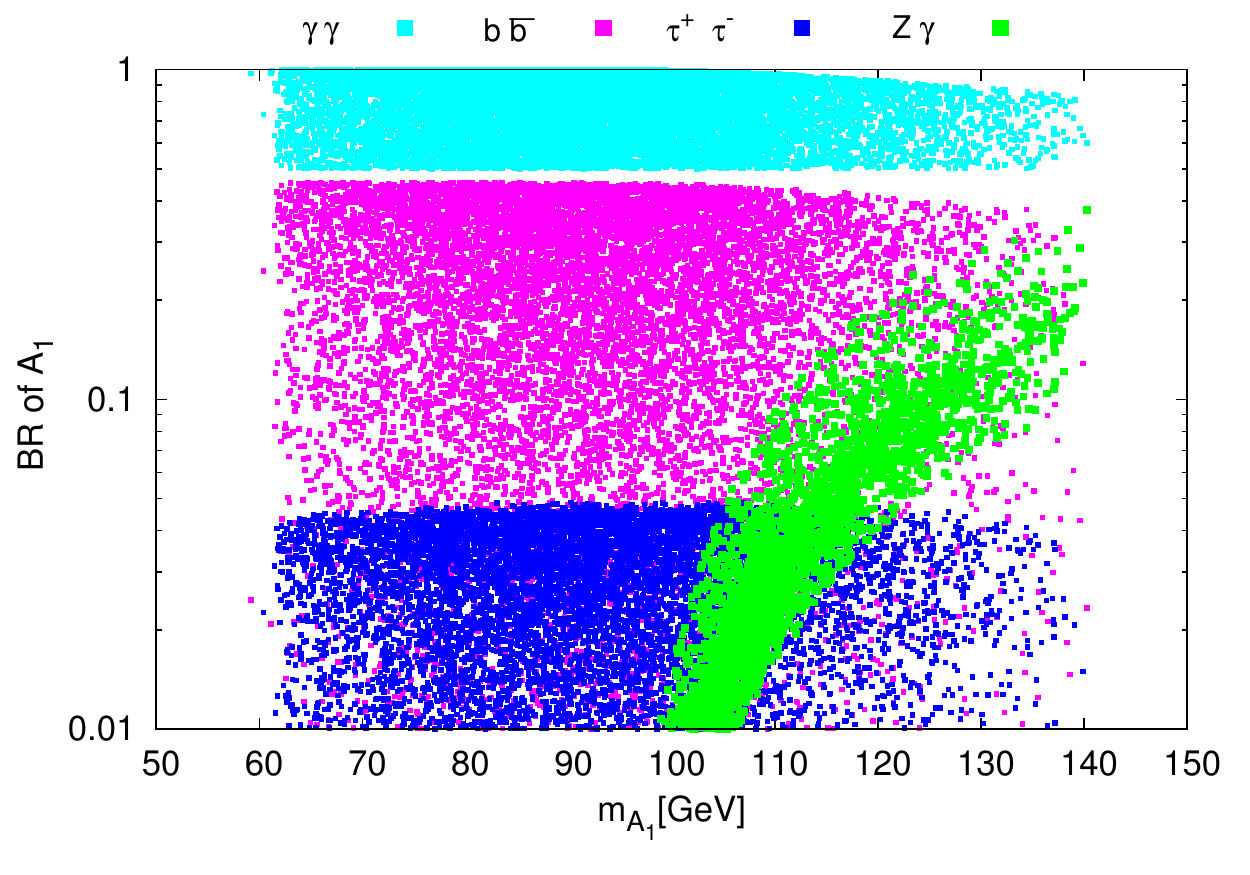}
\includegraphics[width=0.45\textwidth]{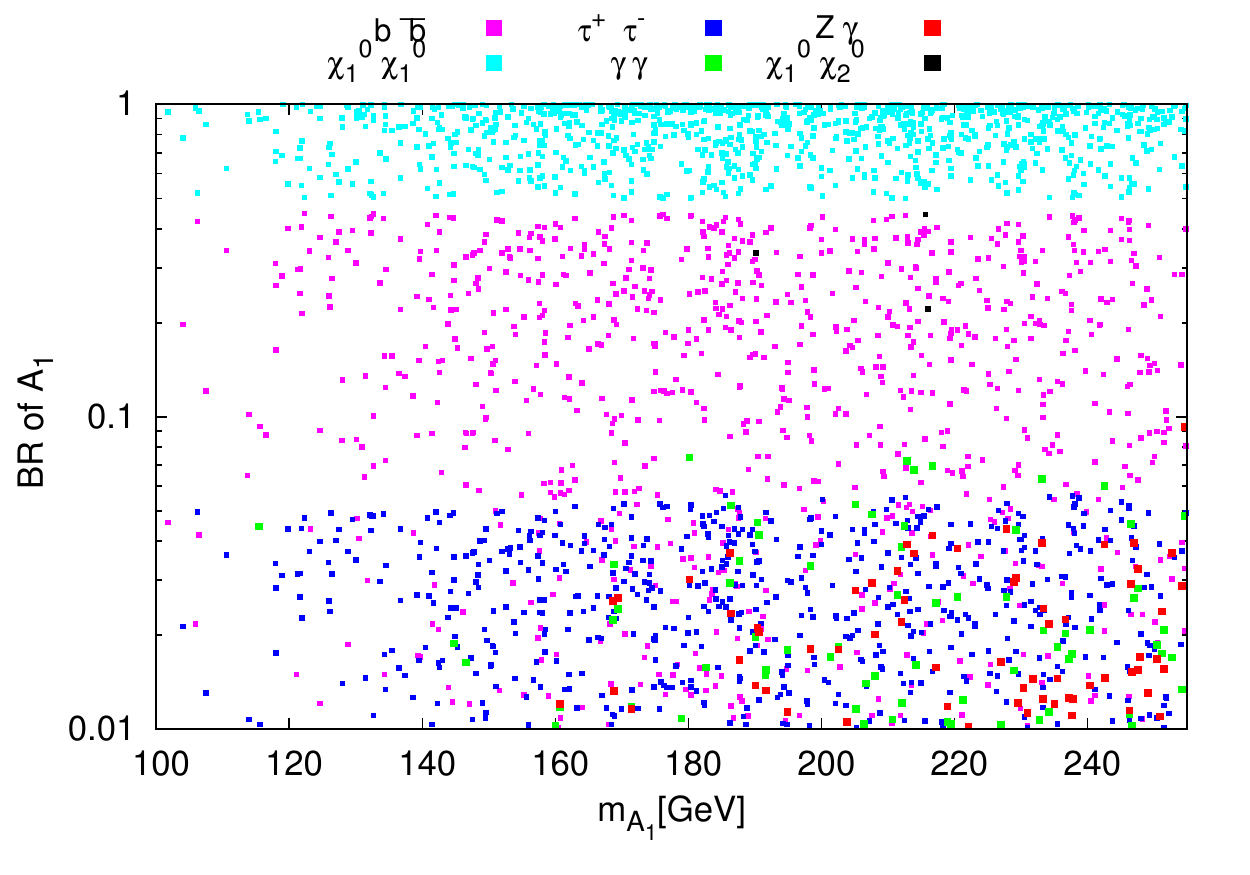}
\caption{Same as in Fig.~\ref{fig:bra1dkh2} but for case(B).}
  \label{fig:bra1dkh1}
\end{figure}

As pointed out earlier, the NMSSM Higgs bosons offer a rich 
phenomenology at colliders owing to their very
diverse decays in various channels including some non SM modes. 
Furthermore, the BR of each Higgs decay modes and hence, the 
sensitivity of the corresponding signal, are very much parameter 
space dependent because of the presence of complicated admixtures 
of physical Higgs states making them either doublet- or singlet-like.
In this work, in order to predict NMSSM Higgs signal at the LHC in the 
context of the scenarios presented in case(A) and case(B),  
we revisit various Higgs decay channels by studying correlations 
among the BRs and the parameter space.  
We also focus on a few interesting Higgs decay 
modes which can lead to robust signals of the NMSSM Higgs bosons, 
in particular, for  
the lightest CP odd and the CP even Higgs bosons. 
In the present context, we discuss features of BRs of various 
decay channels only for two CP even  
($\rm H_1, \rm H_2$) and the  lightest CP odd ($\rm A_1$) Higgs bosons.
\\
$\bullet$ $\rm H_1$ decays : As mentioned before, 
for certain regions of the parameter space in the scenario represented 
by case(A) with $\rm H_2$ as the SM-like Higgs boson, 
$\rm H_1$ and $\rm A_1$ can be very light and can have masses 
below 50 GeV, see Fig \ref{fig:mh1ma1}.    
The decay $\rm H_1 \to \rm A_1 \rm A_1$ will be the 
dominant one whenever it is kinematically allowed.
The corresponding decay width is primarily determined by triple Higgs 
$\rm H_1-\rm A_1-\rm A_1$ coupling mediated by 
singlet components and approximately proportional to $\kappa v_s$, 
thus making the decay width very large in comparison to other 
fermionic and bosonic channels. 
It is found that for certain set of parameters, the BR 
($\rm{H_1 \to A_1 A_1}$) goes to 90\% or more.  
As expected, this triple Higgs boson coupling receives 
corrections~\cite{Nhung:2013lpa, Muhlleitner:2015dua} which are not 
taken into account in this 
present calculation.
There exists a subset of parameter space where $\rm H_1$ decays 
dominantly to $b \bar b$ channel with BR close to 90\% and rest to
the $\tau\tau$ channel ($\sim$ 10\%) as shown in Fig.~\ref{fig:brh1dkh2}(left)
for various masses of $H_1$.
Although the $\rm H_1$ state is dominated by the singlet component,  
presence of a finite fraction of doublet component 
($\rm {S_{i2}}$, composition of $\rm {H_d}$) leads to 
a sizeable coupling of $H_1$ to d-type 
fermions (Eq.\ref{eq:hcoup}).
As a consequence, the partial widths 
of $\rm{H_1 \rightarrow b \bar b,\tau\tau}$ 
modes turn out to be dominant leading to enhancements of respective BRs. 
{
On the other hand, for certain region of the parameter 
space (where $m_{H_1}\gsim$ 60 GeV), the coupling 
$C_{H_1 u \bar u}$ becomes finite (see Fig.2)
resulting in an enhancement of the decay width of $H_1 \rightarrow gg$
(via top quark), $c\bar c$ channels leading a little 
suppression of $b\bar b$  decay BR as demonstrated in 
Fig.~\ref{fig:brh1dkh2} (right),
provided Higgs to Higgs decays are kinematically forbidden.
}Notice that for the mass range around $\rm \mh\sim$ 60 GeV or above, 
the gg, $c\bar c$ channels {are quite large}~\cite{Badziak:2013bda}.
A finite fraction of about 4-6\% of $H_1$ decays to $WW$ for 
higher $\rm \mh$ and also $\sim$2-3\% BR is found for the $\gamma\gamma$ 
channel~\cite{Belanger:2014roa}. 
{ Note that the pattern of BR as shown in 
Fig.~\ref{fig:brh1dkh2}(left and right) correspond to two different
regions of the parameters space.}  
For case(B), $\rm H_1$ is SM-like and 
decays via various SM channels with almost the same rate as in the SM. 
In addition, it can also decay to some non SM modes, 
such as 
$\rm {H_1 \rightarrow \rm A_1 \rm A_1, \tilde\chi_1^0\tilde\chi_1^0}$   
whenever kinematically allowed. Note that,
however, as mentioned earlier, the total 
BR to these non SM channels of SM-like $\rm H_1$ is restricted 
by an upper bound, as given by Eq.~\ref{eq:brns}. Perhaps, looking for 
these non SM decay channels of the SM-like Higgs can be a potential avenue to 
confirm the existence of this type of SUSY scenario.  
\\
$\bullet$ $\rm A_1$ decays : 
For the scenario presented by case (A), 
the lightest CP odd Higgs boson $\rm A_1$ 
primarily decays to fermions due to the absence of tree level 
couplings with gauge bosons. 
As shown in Fig.\ref{fig:bra1dkh2}, for lower mass range, it 
decays dominantly to $\rm b \bar b$ with BR $\sim$90\% 
and then to sub-dominant $\rm \tau\tau$ channels.  
However, if decay modes such as, 
$\rm {A_1 \to \tilde\chi_1^0\tilde\chi_1^0, \tilde\chi_1^0 
\tilde\chi_2^0}$ are accessible 
for a heavier $A_1$ ($\gsim$150 GeV), then those turn out to be
the dominant ones due to the enhancement of $\rm A_1$ couplings 
with singlino-Higgsino
components in $\rm \tilde\chi_1^0$, $\rm \tilde\chi_2^0$ states 
leading to larger width. Here, $\rm \tilde \chi_1^0$ 
and $\rm \tilde \chi_2^0$ are the lightest and the second 
lightest neutralinos respectively.
In this scenario, the 
BR of $\rm A_1$ in the di-muon final state  is very tiny ($\sim 10^{-4}$) 
as shown in the same figure along with other sub dominant decay modes
such as, $\rm {A_1 \to \gamma\gamma, gg, Z\gamma, ZH_1}$.
For the scenario where $\rm H_1$ is SM-like i.e.in case(B), 
the BRs of $A_1$ to various final states are 
depicted in Fig.~\ref{fig:bra1dkh1}. 
Interestingly, 
note that BR($\rm{A_1 \to \gamma\gamma}$) can indeed be very
large ($\sim$100\%) and even greater than the BR for $b\bar b$ channel,
in certain regions of the parameter space and for a wide range of
$\ma$(left). 
This interesting decay pattern of $\rm A_1$ in the di-photon 
channel can be understood by more careful investigation of the
structures of $A_1$ couplings to various particles.     

\begin{figure}[H]
 \centering
  \includegraphics[width=0.45\textwidth]{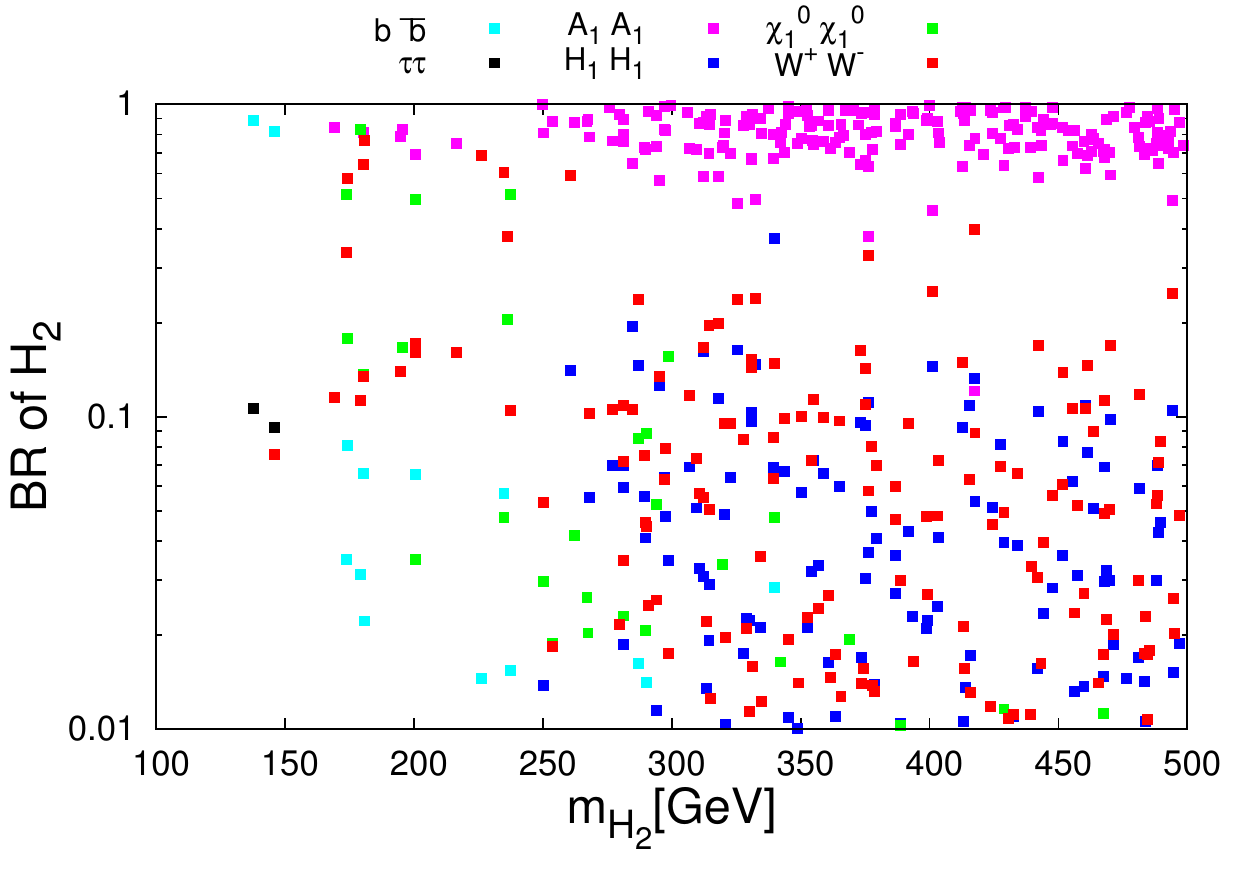}
 \includegraphics[width=0.45\textwidth]{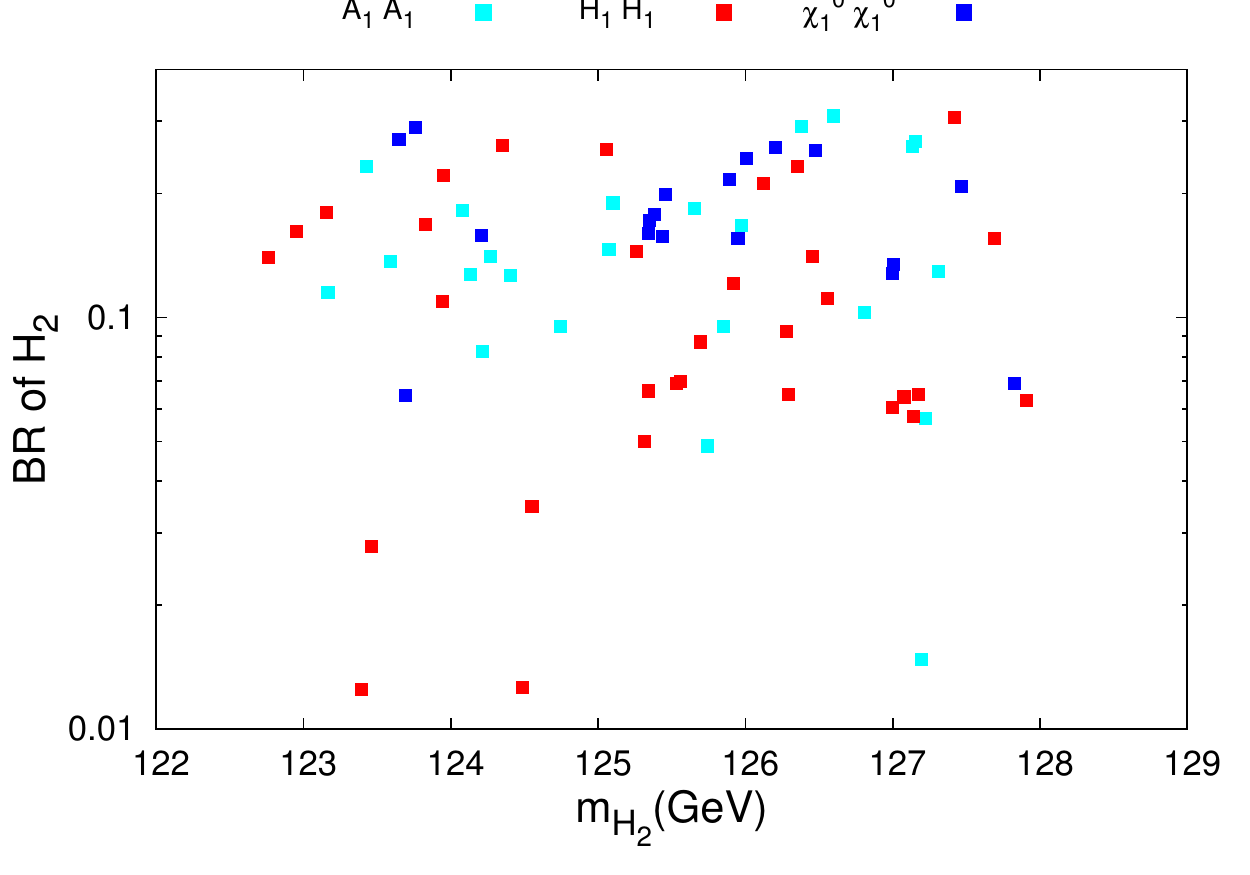}
\caption{BR of $\rm H_2$ for case(B)(left) and case(A)(right) 
subject to all constraints as in Fig~ \ref{fig:brh1dkh2}.}
 \label{fig:brh2h2sm}
\end{figure}

Here, note that $\rm A_1$ is dominantly singlet like resulting in 
all of its available fermionic decay channels suppressed. 
{
Primarily the off-diagonal elements in the Higgs mass matrix
determine the composition of Higgs boson states.
For example, in case of $A_1$
when the off-diagonal term of pseudoscaler Higgs mass matrix vanishes
for a certain combination of the parameters, 
then $A_1$ becomes completely singlet like.
From the 2$\times$2 Higgs CP odd mass matrix, one can conclude that
the vanishing of off-diagonal term leads $A_1$ purely singlet 
dominated which can be translated to the relation such as
$A_\lambda \sim \frac{2\kappa\mu_{eff}}{\lambda}$ and it occurs
for a wide region of parameters.}
Moreover, 
$\rm A_1$ does not have any tree level coupling with the vector bosons. In 
this kind of a scenario, $\rm A_1$ can still couple to photons at 
one loop with charginos in the loop. The Higgsino like charginos favourably 
couple to a singlet like $\rm A_1$ state and enhance the partial width
of  $\rm {A_1 \to \gamma\gamma}$ mode. 
Therefore, for the regions of the parameter space  
where both the charginos are Higgsino like,  
$\rm BR ({A_1 \rightarrow \gamma \gamma)}$ turns out to be very large for a
wide range of $\rm m_{A_1}$ as clearly seen
in Fig.~\ref{fig:bra1dkh1}~\cite{Ellwanger:2010nf,Munir:2013wka}. 
Evidently, this di-photon final state 
appears to be one of the striking features of this model 
and can be exploited not only to discover $\rm A_1$ at the LHC, 
but also to discriminate the MSSM from the NMSSM. It will be discussed 
later in more detail.  
In addition, $\rm {A_1 \to \tau\tau}$
mode is also present with a reasonable BR($\sim$ 10\%) which can also provide a
clean signal of $\rm A_1$. 
{
The other sub dominant decay modes of $A_1$ are presented 
in Fig.~\ref{fig:bra1dkh1} (right) where the dominant one is 
the invisible decay channel, $A_1 \rightarrow \tilde\chi_1^0\tilde\chi_1^0$.  
}
The BR of $\rm A_1$ in the 
muonic channel, $\rm{ A_1 \rightarrow \rm \mu \rm\mu}$ is of the 
same order ($\sim 10^{-4}$) as before along with other sub-dominant 
channels like $\rm {A_1 \to gg,\rm Z\gamma}$. 
{ Note that, both the plots in 
Figs.~\ref{fig:bra1dkh2} and ~\ref{fig:bra1dkh1} 
presenting the BRs of $A_1$ and $H_1$ respectively 
correspond to two different regions of the parameter space for each cases.}   
\\
$\bullet$ $\rm H_2$ decays: In the scenario described in case(B), 
$\rm H_2$ can decay to a pair of 
SM-like $\rm H_1$ as well as to a singlet like $\rm A_1$ pair i.e. 
$\rm H_2 \rightarrow \rm  H_1 \rm H_1, \rm A_1 \rm A_1$. If  
kinematically allowed, these are expected to be dominant($\sim$ 80-100\%) 
is clearly shown in Fig.~\ref{fig:brh2h2sm} (left). 
Since, $\rm H_1$ is SM-like, 
the decay channel, $\rm H_2 \to H_1 H_1$
is likely to contribute to the rate of SM di-Higgs production at the LHC.
Similarly, in Fig.\ref{fig:brh2h2sm} (right) we present BRs of various 
decay channels of $\rm H_2$ for case(A).

In this case, notice that for a good fraction of parameter space 
the mass of $\rm H_2 \sim \rm H_{SM}$ is 
heavier than that of $H_1$ and $A_1$(see Fig 1), 
and BR($\rm {H_2 \rightarrow \rm H_1 \rm  H_1 }$) is found to vary 
between a few percent to $\sim$40\% where as 
BR($H_2 \to \rm A_1 \rm A_1$) is enhanced to 30\%.
Note that, as before, these  are the non SM decay modes of the  
SM-like Higgs $\rm H_2$ and also allowed by the 
constraint in Eq.~\ref{eq:brns}.    
However, the decay channel $\rm H_2 \to H_1 H_1$ is 
expected to contribute to the visible signal of the SM-like 
Higgs boson production. Hence, it is worth  examining  
the contribution of this decay channel in the present Higgs data 
collected at LHC Run I experiments, which eventually may constrain 
the model~\cite{Zhang:2015cta}. 
Notice that in this scenario, case(A), $\rm H_2$ has a 
reasonable BR ($\sim$10-90\%) to a pair of lightest neutralinos 
contributing to its invisible decay width \cite{Ananthanarayan:2015fwa}.
In addition, $\rm H_2$ has many other sub dominant decay modes which 
can provide interesting signals, 
in particular for the scenario 
(B). For example, decays to 
SUSY particles such as a pair of neutralinos, 
${\rm H_2 \to \rm \tilde\chi_2^0 \rm \tilde\chi_2^0,\rm \tilde\chi_2^0 \rm \tilde\chi_3^0}$, may 
provide clean signals because of 
leptonic decays of the neutralinos.
We found that in this regions of the parameter space, 
$ \rm H_2  \to  \rm Z \rm A_1$ is very tiny.  

\section{Higgs production at the LHC}
In general, the NMSSM Higgs bosons can be produced 
via conventional mechanisms, i.e by the dominant gluon-gluon fusion (ggF),
whereas the next sub dominant mode is 
the associated productions with a pair of b quarks. 
{However for a very large value of 
$\tan\beta$, the associated production mode appears to be the 
dominant one.}
We reiterate here that in the NMSSM, the Higgs couplings to  
fermions and the gauge bosons are essentially the SM couplings scaled by 
the respective reduced 
coupling factors which can vary widely from small ($\ll 1$) to moderately
large values($\sim$ 0.5). It implies that the Higgs boson production 
cross-section in the NMSSM can be 
obtained from the corresponding SM cross-section by 
appropriately rescaling them by 
the respective factors. Hence the following naive 
strategy can be adopted to estimate the NMSSM Higgs boson production 
cross section as~\cite{King:2014xwa,Bomark:2014gya},
\br
\rm {\sigma_{ggF}(\phi)} = \rm{\sigma_{ggF}^{SM}(\phi) \times C_{gg}^2}, 
\label{eq:phirate}
\er 
where $\phi$ stands for any of the CP even or odd Higgs boson states. 
Here $\rm{\sigma_{ggF}(\phi)}$ is the NMSSM Higgs boson production 
cross section 
via gluon-gluon fusion and $\rm{\sigma_{ggF}^{SM}(\phi)}$ 
is the corresponding SM cross section for the same Higgs boson mass.
$\rm{C_{gg}}$ is the effective reduced coupling $\rm{gg\phi}$  
via loops comprising heavier { quarks and squarks.}   
The SM Higgs boson production cross-section through ggF is known 
at the level of next-to-next-to leading 
order(NNLO) QCD~\cite{Harlander:2000mg,Catani:2001ic,Harlander:2001is,
Harlander:2002wh,Anastasiou:2002yz,Ravindran:2003um,Blumlein:2005im}.
We compute these SM Higgs boson production cross section via ggF
using SusHi-v1.5.0 ~\cite{Harlander:2012pb} which takes into account 
the QCD contribution up to NNLO mediated by only the SM particles.

However, it is to be noted that in SUSY model, the ggF channel 
may receive a sizeable contribution from loops involving 
the corresponding superpartners, in particular, the third generation 
squarks, the stops and the sbottoms~\cite{Djouadi:2005gj}.
Recently, in the context of the NMSSM, 
the neutral Higgs boson production cross-section via ggF is computed 
including NLO contributions of the squarks and the gluino 
in addition to the 
electroweak corrections involving light quarks and 
it is implemented in SusHi~\cite{Liebler:2015bka}.
These calculations are based on the expansion of terms 
of inverse of heavy SUSY particle 
masses~\cite{Degrassi:2010eu,Degrassi:2011vq,Degrassi:2012vt}.
Obviously, numerical estimation using this calculation 
is expected to be
more reliable than the approximate method 
described in Eq.~\ref{eq:phirate}.
Hence, in this work we use this calculation implemented in 
SusHi-v1.5.0~\cite{Harlander:2012pb} to compute the neutral Higgs 
boson production cross sections in the NMSSM.
Thus, the production rate of Higgs decaying to any of the pair 
of state X can be estimated as, 
\br
\rm R_{XX}= \sigma_{ggF}(\phi) 
\times BR( \phi \rightarrow XX),
\label{eq:phixx}
\er
where $\rm{\phi \rightarrow XX}$ collectively represents the Higgs decay 
channels with 
XX = $\rm{b \bar b, 
\tau \tau, \gamma \gamma, gg, c \bar c, \mu \mu}$,
WW, ZZ  etc. 
For a given decay channel and a set of parameter values,  
$\rm{BR(\phi \rightarrow XX)}$ is obtained from the NMSSMTools.
 $\rm{\sigma_{ggF}(\phi)}$ is computed 
directly using the NLO calculation~\cite{Liebler:2015bka} in the NMSSM 
implemented in SusHi.
On the other hand, for the sake of comparison, we also estimate
the cross section for the same Higgs boson production with the same
mass following the approximate method as described in 
Eq.~\ref{eq:phirate}. It is found that following this prescription 
the calculated CP even Higgs boson 
production cross sections are 30-40\% higher than the value 
obtained using a more exact calculation of SusHi~\cite{Liebler:2015bka}.
This can be attributed to the fact that in the NMSSM, the production cross 
sections are estimated at NLO including electroweak corrections, 
where as in Eq.~\ref{eq:phirate} NNLO QCD result without any SUSY loops 
are used along with $\rm C_{gg}$ estimated at NLO.

The cross-section for the next-dominant production mode, can be 
obtained by exclusive Higgs boson production in association with a pair
of b quarks~\cite{Dittmaier:2011ti},
\br
\rm{gg \rightarrow b \bar b \phi}
\er 
However, this cross section can be obtained also by computing 
$\rm{b \bar b \to \phi}$ which is a good approximation within the higher 
order correction~\cite{Dittmaier:2003ej,Maltoni:2003pn}.
Note that the SusHi calculates this $\rm{b \bar b \to \phi}$ 
at NNLO QCD. The rate 
$\rm R_{XX}$ corresponding to this production channel is obtained 
by multiplying the effective 
reduced $\rm b \bar b \phi $ coupling factor with the 
SM production cross section obtained from SuSHi\cite{Harlander:2012pb} 
and the respective BR($\phi \to XX$). 
In our calculation we use the CT10~\cite{Gao:2013xoa} for parton
distributions function 
and both the QCD scales are set to $\rm M_{\phi}$ for ggF, whereas  
the factorization scale is fixed to $\rm {0.25 M_{\phi}}$ 
only for $\rm{b\bar b \rightarrow \phi}$ production. 
Needless to say that the rate of the Higgs production in any of the decay 
channels is very much sensitive to the model parameters  
owing to the presence of mixing angles in the Higgs couplings, 
see Eq.~\ref{eq:hcoup} and \ref{eq:acoup}.

Indirectly, $\rm H_1$ and $\rm A_1$ can also be indirectly produced in pairs 
through $ \rm H_2$ production via ggF and its subsequent decays 
provided $\rm m_{H_1},m_{A_1} <  \mhh/2$. 
The corresponding production rate can be obtained as,
\br
\rm \sigma_{ggF}^{\phi\phi}(H_2) &=& \rm {\sigma_{ggF}(H_2) 
\times BR(H_2 \to \phi \phi)} 
\er
\label{eq:rbar}
where $\phi=\rm H_1,\rm A_1$. As before, the $\rm{\sigma_{ggF}(\rm H_2)}$ 
is computed 
from NLO calculation in the 
NMSSM framework using SusHi~\cite{Liebler:2015bka}.

Now we present the light neutral non-SM Higgs bosons production 
rates $\rm R_{XX}$
(Eq.~\ref{eq:phixx}) via both the production modes, 
ggF and $\rm{b \bar b \to \phi}$ 
for various decay modes corresponding to the regions of 
parameter space presented by case(A) and case(B). 
As pointed out earlier, the opening up of many 
decay modes of the non-SM-like Higgs bosons offers a variety of signals with 
a widely ranging rates. In order to demonstrate, we compute the 
minimum and the maximum values of those rates, $\rm R_{XX}$ for those   
region of the parameter space for a given mass of the Higgs boson 
and the decay channel. Naively, such regions of the parameter 
space which leads to the 
minimum and maximum cross sections 
for a given Higgs mass and decay channel 
are identified by the corresponding values of the product,    
$\rm{C_{gg}^2 \times BR(\phi \to XX)}$. We assume that the 
parameter region 
which gives the maximum(minimum) value of this product yield
a maximum(minimum) value of the cross section for the 
given Higgs boson mass and the decay channel.   
\begin{table}[ht]
\centering
\footnotesize
{\renewcommand{\arraystretch}{1.9}
\begin{tabular}{|c|ccccccc|}
\hline
Rates & & & &$\mh$(GeV) & & & \\
& 13   & 28 &  48    &   68  &   89  & 110  &   120 \\ \hline
$R_{b \bar b}(gg)$ & 144927 & 2383 -- 37088 & 1164 -- 9835 & 126 --4735 
& 4.5 -- 4345 & 17.4 -- 2337  & 45 -- 1126 \\
$R_{b \bar b}(b\bar b)$  &  1773 &189 -- 2952 &170-- 2133 & 1868 & 868 
& 421 &6 -- 249 \\
\hline
$R_{c\bar c}(gg)$ & 423 &10 -- 46 & 1.3 -- 60 &0.03 -- 3737& 9E-4 -- 1726 
& 4.5E-3 -- 808 &0.07 -- 11 \\
$R_{c\bar c}(b \bar b)$ & 5.18 &0.8 -- 3.15 &0.4 -- 4.4 & 4.3 & 3.2 
&3E-4 -- 1.5 &3E-2 -- 1.1 \\
\hline
$R_{\tau\tau}(gg)$ & 13838 &  213 -- 2961 &105 -- 858 &5E-3 -- 455 
& 1.3E-2 -- 448 &6.4E-2 -- 250& 5--120 \\
$R_{\tau\tau}(b \bar b)$  
& 169 & 17 -- 236 & 16 -- 186 & 180 & 87 & 44 & 0.61 -- 27 \\
\hline
$R_{\mu\mu}(gg)$ & 55 &0.76 -- 11 &0.37 -- 3 & 1.6& 1.6 &2.2E-4 -- 0.88 
&1.7E-2   -- 0.42 \\
$R_{\mu\mu}(b \bar b)$  & 0.7 & 0.06 -- 0.85 &5.6E-2 -- 0.66 & 0.64 
& 0.3 & 0.15 &2.2E-3 -- 9.6 \\
\hline
$R_{\gamma \gamma}(gg)$ & 1.4 & 0.01 -- 0.47&2.6E-3 -- 1&2.6E-4 -- 93
& 79&2E-4 -- 58& 2.5E-4 --0.88 \\
 $R_{\gamma \gamma}(b \bar b)$  & 1.7E-2 &1E-3 --  3.2E-2 &7E-4 -- 0.09 & 0.17 & 0.22 & 0.18 &1.2E-4 -- 0.1\\
\hline
$R_{gg}(gg) $ &1028 &12 -- 161  & 3 -- 38 &0.22 -- 2881 &5.6E-3 -- 2077 
&0.02 -- 1351 & 0.11 -- 15   \\
$R_{gg}(b \bar b) $   &12.6 & 0.93 -- 12.8 & 0.8 -- 6 & 4.8 & 1E-4 -- 4.5 
&4.3E-4 -- 3 &4.3E-2 -- 1.6 \\
\hline
\end{tabular}
\caption{Production rates(cross section $\times$ branching ratio)( in fb)
of $\rm H_1$ at 13 TeV LHC energy for case(A).}
\label{tab:H2smRH1}
}
\end{table}
In Tables~\ref{tab:H2smRH1} -- \ref{tab:H1smRA1}, we present minimum and 
maximum rates for non-SM-like Higgs bosons in various 
decay channels for the scenarios case(A) and case(B).
For example, in Table~\ref{tab:H2smRH1}, the results are shown for the 
non-SM-like 
$\rm H_1$ for case(A).  
Rates are presented for the Higgs production via both  
ggF ($R_{XX}(gg)$) and associated production, 
$b \bar b \to \phi$($R_{XX}(b \bar b)$).
In this table, the first row shows the selected values of $\rm \mh$ extending 
up to 120 GeV, while the first column indicates the 
rates ($\rm R_{XX}$) for the decay channel XX 
of which minimum and maximum values are presented. 
For instance, with a choice of lower value of $\rm{\mh=13}$ GeV, 
the maximum rate of $\rm H_1$ production in $b\bar b$ channel via ggF 
is about 145$\times 10^3$~fb, where as the same for $\gamma\gamma$ and 
$\tau\tau$ channels are about 1.4 fb and 14 $\times 10^3$fb, respectively. 
As seen in this Table that the 
rates($\rm R_{XX}(b\bar b)$) 
via associated production Higgs are also not negligible. 
Note that we do not present the minimum values of 
the cross sections if it is found to be less than $10^{-4}$ fb.
Also, we only present single numbers for those cases where we have very few
allowed points in the parameter space and no presentable variation in rates
is observed for them. 
It is well known that the signals in the  
$\gamma\gamma, \tau\tau$ channels are comparatively clean due to 
the efficient tagging of photon and improved detection of $\tau$ leptons
either in jets or leptonic final states. For higher values of 
$\rm \mh$, say for 110 GeV, the dominant cross section is found to 
be again for the $b\bar b$ final state 
values ranging over 
from a few 
fb to a few thousands fb,  whereas in the case of the $\tau\tau$ final state, 
the maximum value lies within a few 100 fb. 
Interestingly, note that the rate of the $\rm H_1$ production decaying 
in the $\rm H_1 \to gg$ channel is reasonably large for all masses, 
unlike the situation in the SM as well as in the MSSM, 
where it is very much suppressed.
Hence this channel can have the potential candidate to distinguish the
NMSSM from other SUSY models, although elimination of QCD background 
is a non-trivial task.    
Remarkably, we observe that
at higher values of $\rm \mh$ for which $\rm{\ma \lsim \mh/2}$, 
Higgs to Higgs decays, such as $\rm H_1 \to \rm A_1 \rm A_1$ open up with a 
maximum rate found to be $\cal O$(1000 fb) which is not shown here.
In this case, of course, the possibility of observing the  
signal of $\rm H_1$ depends on $\rm A_1$ decay pattern and the 
measurement can provide the information about triple Higgs bosons
coupling.
It is also observed that the rate in the $\rm WW$ channel at the higher 
masses range($\sim$ 120 GeV) is non negligible, and can vary between  
the range of ${\cal O}$(1 - 100~fb).  {{In addition, possibility
of the $\rm{H_1 \rightarrow  \tilde\chi_1^o \tilde\chi_1^0}$   
(invisible decay) exists for higher $\rm \mh$
($\gsim$100 GeV) with a very tiny rate in a certain region of 
parameter space.} }

\begin{table}[!htb]
\centering
\footnotesize
{\renewcommand{\arraystretch}{1.9}
\begin{tabular}{|c|cccccccc|}
\hline
Rates &  & &  &&$m_{A_1}(GeV)$ & & &\\
&22 & 30 & 60 &100 &140 &180 &218 &249    \\
\hline
$R_{b \bar b}(gg)$  & 72336 & 36759 &818 -- 1685 &143 -- 226 &1.53 -- 29 
&0.08 -- 0.3 & 0.018 -- 0.09 &9E-3 -- 0.6  \\
$R_{b \bar b}(b \bar b)$ & 1392 & 2274 & 385 -- 869 & 115 -- 262 & 4 -- 65 
& 0.3 -- 1.1 & 0.06 -- 0.18 &
0.04 -- 0.06 \\
\hline
$ R_{c \bar c}(gg)$ &  62 & 26 & 0.3 -- 0.6 &1.7E-2 --  0.04 &1.6E-4 -- 3.5E-3
& 3.5E-5 & - & 2.1E-3\\
$ R_{c \bar c}(b \bar b)$  & 1.19 & 1.57 & 0.13 -- 0.3 & 0.02 -- 0.05 
& 4E-3 --7E-3 & -- & -- &1E-3 \\
\hline
$R_{\tau \tau}(gg)$ & 5939 & 2813  &76 -- 156 &14 -- 23 
&0.17 -- 3.2&9E-3 -- 3.4E-2 
& 2.2E-3 -- 1.1E-2 &1E-3 -- 7E-2   \\
$R_{\tau \tau}(b \bar b)$ & 99 & 174 & 36 -- 82 & 12 -- 27 & 0.5 -- 7 
& 0.034 -- 0.12 & 0.007 -- 0.02 &
5E-3 --7E-3 \\
\hline
$R_{\mu \mu}(gg)$ & 18.4&10 &0.3 -- 0.55 &0.05 -- 0.08
&6E-4 -- 1.2E-2& 1.1E-4& -
&  2.4E-4  \\
$R_{\mu \mu}(b \bar b)$  & 0.35 &0.62 & 0.12 -- 0.3 & 0.04 -- 0.1 
& 0.002 -- 0.02 & 1E-4 -- 4E-4 & - &- \\
\hline
$R_{\gamma \gamma}(gg)$ & 0.9& 0.74 & 0.08 -- 0.15 &0.04 -- 0.07 
&1E-3 -- 1.7E-2 
&1E-4 -- 4E-4& 1.4E-4 & 3.5E-3  \\
$R_{\gamma \gamma}(b \bar b)$  & 0.02 &0.05 & 0.04 -- 0.1 & 0.03 -- 0.09 & 
0.003 -- 0.04 & 3E-4 -- 14E-4  & 2E-4 -- 3E-4 
&(E-4 -- 2E-4 \\
\hline
$R_{gg}(gg)$&  465 & 186.2 &1.8 --  4.1 &0.14 -- 0.3 &1E-3 -- 2.2E-2& 1.1E-4& -
& 1.2E-2 \\
$R_{gg}(b \bar b)$ & 9 &11.5 & 0.9 -- 1.8 & 0.14 -- 0.31 & 0.003 -- 0.05 & (1--4)E-4 & - &8E-4 \\
\hline
\end{tabular}
\caption{ Production rates(cross section $\times$ branching ratio)(fb)
of $A_1$ at 13 TeV LHC energy for case(A).}
\label{tab:H2smRA1}
}
\end{table} 
Similarly in Table~\ref{tab:H2smRA1}, we present the minimum and 
the maximum 
rates of $\rm A_1$ in any given decay channel through both the 
production modes for a range of masses {{22 - 249 GeV}} for the scenario 
represented by the case(A).  
The dominant rates 
of $\rm A_1$ also appear to be in the $b \bar b$ decay channel. On the 
other hand, $\tau\tau, \gamma\gamma$ channels predict a range of 
sizeable rates, in particular
for $\rm {m_{A_1}} \sim$ 100 GeV or less. 
For higher 
masses of $\rm A_1$, the rates for $\tau\tau, b \bar b$ channels 
are reasonable.  
Decays  of $\rm A_1$ in various SUSY particles,  
in particular, for higher $A_1$ masses also appear with a few fb cross 
section which are not presented here. 
\begin{table}[ht]
\centering
\footnotesize
{\renewcommand{\arraystretch}{1.9}
\begin{tabular}{|c|cccccccc|}
\hline
Rates &  & & &$m_{A_1}(GeV)$ && & & \\
&  17  & 30  & 59  &  100     &  138     &  181  & 220    & 249   \\ \hline
$R_{b \bar b}(gg)$ & 36253 & 10164 &.01 -- 686 &6.3E-4 -- 99&4.5E-4 -- 29& 7 
& 0.55& 4.8   \\
$R_{b \bar b}(b \bar b)$  & 202 & 1058 & 0.004 -- 263 & 0.001 -- 123 & 0.001 -- 33 
& 28 & 1.27 &1.5 \\
\hline
$R_{c \bar c}(gg)$ & 39 & 6.4 & 0.25&  0.45& 0.07 & 0.022& 1.2E-3 & 0.18 \\
$R_{c \bar c}(b \bar b)$ & 0.21& 0.66 & 0.1 & 0.03 & 0.008 & 0.002 & 0.0001 & 0.0013 \\
\hline
$R_{\tau \tau}(gg)$ & 2421 & 833  &.001 -- 62 & 10 & 3.12 & 0.81 & 0.07 & 0.58   \\
$R_{\tau \tau}(b \bar b)$ & 13.4 & 87 & 4E-4 -- 24 & 1E-4 -- 13.6 & 1E-4 -- 3.7 
&  3.2 & .15&.184 \\
\hline
$R_{\mu\mu}(gg)$  & 8.8& 3.0 & 0.22  & 0.035 & .011 &2.8E-3 & 2.5E-4& 2.1E-3 \\
$R_{\mu\mu}(b \bar b)$ & 0.05& 0.31 & 0.1 & 0.05& 0.013 & 0.011 & 5E-4 & 6E-4\\
\hline
$R_{\gamma \gamma}(gg)$  & 0.3& 0.2 & 4E-3 -- 0.11  & 0.042   & 0.058 & 0.017 
& 8.4E-3 & 0.035 \\
$R_{\gamma \gamma}(b \bar b)$& 2E-3 &0.01  & 2E-3 -- 0.058 & 1E-4 -- 0.06 
& 0.04 & 6E-3 & 3E-4 & 2.1E-3 \\
\hline
$R_{gg}(gg)$ & 305 &46.3  & 2.5E-3 -- 1.7   & 0.95 & 0.22& 0.1 & 6E-3 & 0.97 \\
$R_{gg}(b\bar b)$ & 1.7 & 4.8  & 0.66 & 0.15 & 0.025 & 0.014 & 7E-4 &7E-3 \\
\hline
\end{tabular}
\caption{
 Production rates(cross section $\times$ branching ratio)(fb)
of $A_1$ in various decay channels
at 13 TeV LHC energy for case (B).}
\label{tab:H1smRA1}
}
\end{table}
In Table~\ref{tab:H1smRA1}, as before, we present the rates of the 
$\rm A_1$ production for a mass range from $17-249$ GeV in multiple 
channels for case(B) where $\rm  H_1$ is the SM-like Higgs.
Note that as before the experimentally clean 
$\rm \gamma \gamma$ and $\rm \tau \tau$ channels provide a 
sizeable rates for lower values of $\rm m_{A_1}$, whereas for  
higher values of $\rm m_{A_1}$, depending on the regions of the 
parameter space, the rates for $\tau\tau, b \bar b$ channels appear to be 
feasible for exploring $A_1$.  The $c\bar c, gg$ channels are 
not very promising due to tiny rates and  
presence of large QCD background, except for the lower masses
where the rates are sizeable.
As observed before for the $\rm H_1$ case, the $b\bar b$ channel has  
the most dominant rates for all masses of $\rm A_1$. 
The production rate of heavier non-SM-like Higgs boson $\rm H_2$ in
various final states are presented in Table~\ref{tab:H1smRH2} for
the case(B). For the lower mass range, just above the
SM-like Higgs boson mass, the dominant rate is due to the $b\bar b$
mode as expected followed by the same in
$\tau\tau$ channel which are not shown.
For higher mass range around $\sim$ 200~GeV, the rates for
the channels such as, WW, $\rm A_1 \rm A_1$ and $\rm H_1 \rm H_1$ are
quite large.  It is also observed the rate in the invisible decay mode
($H_2 \to \tilde\chi_1^0\tilde\chi_1^0$) is not negligible in a
certain region of parameter space. Furthermore, for
very high values ($\mhh \ge $200~GeV), Higgs to Higgs decay modes
appear to be the dominant ones and provide
a moderate rates ranging between a few fb to a few tens of a fb.

\begin{table}[H]
\centering
\footnotesize
{\renewcommand{\arraystretch}{1.9}
\begin{tabular}{|c|cccc|}
\hline
Rates &  & &$m_{H_2}$(GeV) &   \\ 
&137& 200&300 &400  \\  \hline
$R_{b \bar b}(gg)$  &32  &1.1 -- 7.68 &2.5E-4 -- 0.18 &1.6E-4 -- 0.05     \\
$R_{b \bar b}(b \bar b)$  &78  &7.4E-3 -- 1.5 & 9.1E-4 -- 0.066 & 
1.5E-3    \\ \hline
$ R_{\tilde \chi_1^0 \tilde \chi_1^0 }(gg)$&  -    
&30 --  58.32& 5.0 & 7E-4 \\
$ R_{\tilde \chi_1^0 \tilde \chi_1^0 }(b\bar b)$ &-  &0.19 -- 11.51 &3.5E-4 -- 1.8 & 1E-4     \\ \hline
$R_{WW}(gg)$  & 0.1&19 -- 149 & 3.1 &1E-3 -- 13  \\
$R_{WW}(b \bar b)$  & 0.24  &0.98 -- 3.7&5.6E-4 -- 0.28 &2.2E-4 -- 0.26     \\ \hline
$R_{A_1 A_1}(gg)$& - & 602 &37 &0.35 -- 24   \\
$R_{A_1 A_1}(b \bar b)$ &- & 4 & 3&4.5E-2 -- 0.95     \\ \hline
$R_{H_1 H_1}(gg)$& 0 & 0 & 2.8&3.6E-4 -- 7.5    \\
$R_{H_1 H_1}(b \bar b)$ &-  & - & 0.4 & 0.15     \\ \hline
\end{tabular}
\caption{
Production rates(cross section $\times$ branching ratio)(fb)
of $\rm H_2$ 
at 13 TeV for case (B)}
\label{tab:H1smRH2}
}
\end{table}

\section{Higgs signal}
The discovery potential of the non-SM-like Higgs boson depends 
on the underlying production mechanism along with its subsequent 
decay spectrum and the resulting rate.
Already many studies exist in the literature 
presenting the detection prospect of finding 
the non SM Higgs bosons in the NMSSM exploiting,
in particular popular $b \bar b,\tau\tau$ modes producing Higgs bosons directly
or indirectly via Higgs to Higgs 
decays (for a review and details see	
Refs.~\cite{Ellwanger:2011sk,Bomark:2014gya,Ellwanger:2004gz} and references
therein). However, here we try to emphasize the detection prospect of 
the non-SM-like Higgs signal via enhanced rate in 
the $\gamma\gamma$ final state and as well as the other 
final states such as gg, cc. 
 
As we have noticed in the previous sections, 
the BR in $\gamma\gamma$ mode of 
$\rm A_1$ and $H_1$ can be very large,  
in particular for $A_1$, it is closer to 
90-100\% in a certain region of parameter space  
in the scenario case(B), see Fig.\ref{fig:bra1dkh1}
\cite{Christensen:2013dra,Ellwanger:2010nf,King:2014xwa,Benbrik:2012rm}
{ and similarly the BR of SM-like Higgs boson of mass 125 GeV 
in the di-photon is also enhanced in comparison to SM Higgs 
boson~\cite{Ellwanger:2011aa}}.   
The BR of $\rm H_1 \to \gamma\gamma$ is also not negligible 
and is observed to be around 2-3\% for $\mh$ masses $\sim$60~GeV or above 
as seen in the Fig.\ref{fig:brh1dkh2} and the 
Table \ref{tab:H2smRH1}~\cite{Jia-Wei:2013eea}.
For example, for $\mh \sim 68$~GeV,
(see Table~\ref{tab:H2smRH1}), at 13 TeV LHC energy, 
one can have maximum about few 
thousands events in $\gamma\gamma$ channel  
for the integrated luminosity 100$\invfb$ 
For the same scenario, in case of $\rm A_1$, for 60 GeV mass, 
one can expect about fewer  
event $\sim$10-15 (see Table~\ref{tab:H2smRA1}) for the same 
integrated luminosity. 
Here, the large BR of $\rm A_1$ into two photons channel does not 
yield a large rate because of suppressed production 
cross section. As is well known, looking for a signal in the di-photon 
final state is very promising as photon is a very clean object 
to tag experimentally. It is remarkable that neither in the SM 
nor in the MSSM, the decay rates of the Higgs bosons in the di-photon 
channel is comparable to what is in the NMSSM.
For instance, in Refs.~\cite{Ellwanger:2010nf,Benbrik:2012rm} it is 
shown that for the $\rm H_1$, the relative signal rates i.e production 
cross sections times BR($\rm H_1 \to \gamma\gamma$) 
are almost larger by almost a factor of 5-7 
what is predicted by the SM for $\rm \mh \lsim 100$GeV {and 
similar pattern is also observed even for $\rm \mh \gsim 100$GeV 
case}~\cite{Cao:2011pg}.
{In this current study we focus on the di-photon channel
for a very low mass region($<$125 GeV) of non SM-like Higgs bosons which
are accessible at the current LHC experiments.
After performing a very systematic scanning of the parameter 
space we present the minimum
and maximum rates of di-photon production for a given mass of non-SM-like Higgs. In addition we also discussed the variation
of the BR of the di-photon channel for this mass range along with other decay 
channels. 
It is worth to have this information 
in order to estimate the sensitivity of non SM-like Higgs production for
a given integrated luminosity.
}
Evidently, this diphoton channel appears to be a smoking gun
signal to look for light neutral non SM Higgs bosons.
Although some { simple parton-level analyses } 
were already carried out in the past to explore  
the detection possibility in this di-photon channel 
~\cite{Cao:2011pg, Moretti:2006sv}, more detail analysis are 
required in the present context. The estimation of the corresponding
SM background in this di-photon channel is also a challenging task, 
in particular, for the low mass Higgs bosons. We postpone this analysis 
to a future work\cite{guchait:2015jk}.

Furthermore, for Higgs decays to two gluons,
$\rm H_1 \to gg$ with a branching ratio of about 50\% or 
more, a reasonable number of events in this channel are expected for the 
100$\invfb$ integrated luminosity (see Fig.~\ref{fig:brh1dkh2} 
and Table~\ref{tab:H2smRH1}), in particular 
for the mass range $\sim$60-100 GeV, this channel  
yields a huge number of events (${\cal O}$(1000)) even for an l 
integrated luminosity as low as 10$\invfb$.
Clearly, this gg final state provides an alternative option to detect 
the NMSSM light neutral Higgs bosons at the LHC by probing the 
invariant mass of a pair of jets, which is not a 
easy task due to the presence of
enormous QCD background. 
Perhaps, in the  $\rm H_1$+1 jet final state,  
where an extra jet comes from radiation followed by  
$\rm H_1 \to gg$ decay, one can tag two jets to reconstruct 
the Higgs mass using the jet substructure method which is a very efficient 
technique~\cite{Altheimer:2012mn} to probe boosted objects.
Of course, the feasibility of 
this channel can be understood only after carrying
out a detailed simulation including QCD background. 
Possibly, this $gg$ channel provides a direct  
opportunity for a direct measurement of effective Higgs coupling to gluons 
directly~\cite{Langenegger:2015lra}. 
In addition, the  
decays of the Higgs bosons to SUSY particles, such as, 
$\rm{\phi \to \tilde\chi_1^0 \tilde\chi_2^0}$, 
if are kienamtically allowed and have a reasonable BR, can also lead to  
interesting signals, particularly for leptonic decays of sparticles. 
Essentially, sparticle decay channels open up for values of 
$\mh \sim$200 GeV. These decay channel  
$\phi \tilde\chi_1^0\tilde\chi_1^0$ may also contribute to the 
invisible width of the Higgs boson. Finally, Higgs to Higgs decay 
channels (Eq.\ref{eq:h12dk}) give rise
to many more detectable final states depending on the decays of $H_1$ and
$A_1$. A detail investigation of these channels and measurements of 
triple Higgs couplings may reveal the model structure of the  
NMSSM.  
 In Table ~\ref{tab:bps}, for the sake of illustration
we present for few cases some of the relevant parameters 
where the rates are reasonably large for a given Higgs boson mass and
decay channel.
\begin{table}[H]
\centering
\footnotesize
{\renewcommand{\arraystretch}{1.5}
\begin{tabular}{ |c|c|c|c|c|c|c|  }
\hline
& P1&P2 &P3& P4 \\ \hline
Case &A& A& B&B \\ \hline
$\rm M_{H_{i}}$ (GeV) &13 (i=1)&67 (i=1)& 200 (i=2)&400 (i=2)  \\
$R_{XX}(gg) (fb)$ &144927$(b \bar b),13838(\tau  \tau)$ &$93(\gamma \gamma),~ 2881(gg),~3737(c\bar c)$&$602(A_1 A_1),~ 149(WW)$&  $24(A_1 A_1),~ 13.03(WW)$  \\ \hline

$\lambda, ~\kappa$&0.363,~0,148 &0.325,~0.167& 0.544,~0.265 & 0.64,~0.317    \\
$ \tan \beta ~\mu$ &9.27~,165.45&8.36~,229.02 &2.45~,192.7 &2.33,~401.30  \\
$A_{\lambda},~ A_{\kappa}$ & 1440.7, -273.95& 1658.38,-450.35&508.06,~3.69  &530.27,-35.35\\
$M_{Q_3},~M_{U_3}$   & 2828.25,~358.58 & 1847.6,~ 2950.2& 2593,~ 1686.8 &1138.32,~ 1427.90     \\
$A_t$   & -2574.43 & -2779.78& -3000.0 &-3016.5     \\
$M_1,~ M_2$&89.18,~335.99 &75.9,~414.18& 92.89,~816.69& 80.6,~ 114.3 \\ \hline
 & P5&P6 &P7& P8 \\ \hline
 Case &A& B& A&B   \\ \hline
$\rm M_{A_{1}}$ (GeV)&22 & 60& 100&249   \\
$R_{XX}(gg) (fb)$& $72336(b \bar b)$, 5939$(\tau \tau)$&$0.112(\gamma \gamma)$& $226(b \bar b)$, 23$(\tau \tau)$ & $0.035(\gamma \gamma)$  \\ \hline

$\lambda, ~\kappa$ &0.368,~0.103& 0.328,~0.434&0.677,~0.182 & 0.499,~0.476    \\
$ \tan \beta ~\mu$ &8.56~,154.74 &5.61~,128.7&7.66~,193.0 &1.55,~1859.0  \\
$A_{\lambda},~ A_{\kappa}$ & 1432.96,-0.486& 553.3, -1.53& 1464.15,-81.22  &872.53,-13.41\\
$M_{Q_3},~M_{U_3}$   & 2237.9,~ 829.6 & 2744.15,~ 1441.90& 1962.45,~ 2666.53 & 1289.96,~1165.72    \\
$A_t,~A_b$   & -2516.9 & -3914.36& 1850.3 &-1617     \\
$M_1,~ M_2$ &396.07,~319.15&98.19,503.6& 410.8,160.43&56.2,~ 574.26  \\
\hline

\end{tabular}
\caption{Parameter sets(P1 - P8) for a given Higgs boson mass and decay channel
for both the cases A and B corresponding to the rates $R_{XX}$(in fb).
These points P1-P8 are selected from Tabe 2-5.}
\label{tab:bps}
}
\end{table}

\section{Summary}
The recently discovered Higgs particle can be accommodated in the framework
of the NMSSM easily without much adjusting the input parameters unlike
in the case of the MSSM. 
Scanning the parameter space taking into account all constraints, we found 
the scenario where one of the Higgs bosons is SM-like and,  
some of the singlet like neutral Higgs bosons can be rather 
very light with suppressed couplings to fermions and the gauge bosons. 
In this work we focus our attention only to light neutral two CP even 
Higgs bosons and one CP odd Higgs boson out of which one the 
CP even Higgs bosons is SM-like. 
We discuss the masses and the couplings of the non-SM-like light neutral 
Higgs bosons and their BRs in various decay channels which predict
the Higgs phenomenology at colliders. We present the BRs of the 
non-SM-like Higgs bosons in
all accessible decay channels in reference to the allowed parameter space.
We present the ranges of over all rates of production 
for a given non SM Higgs mass and for a given decay channel
in the allowed region of parameter space, which demonstrate the 
potential final states that can be probed in the search for 
non SM signal at the LHC. 
We found some of the characteristic decay channels which are less dominant
in the the SM and in the MSSM could play 
important roles in such searches. 
For example,
in certain region of the parameter space, a large BR of Higgs decay to 
two photons is found to be quite interesting and 
can provide not only a characteristic robust signal of the NMSSM, but 
can also help to discriminate the NMSSM from the MSSM. 
{ Although this observation is also made in previous analyses, as discussed
in sec. 5, but here we present the rates in this di-photon channel 
more systematic way for various lower masses of Higgs bosons 
corresponding to the center of mass energy at 13 TeV. These information 
may be very useful in searching for the non SM-like Higgs bosons at the LHC.}
Higgs decay to two gluons also turn out to be an interesting 
mode which  one can exploit to study the
the NMSSM Higgs sector, though care must be taken to deal with an 
enormous QCD background.
Moreover, further studies of Higgs to Higgs decays via triple Higgs 
bosons couplings and probing the associated signals in
various final states could uncover the underlying dynamics of the model.

\section{Acknowledgement}
The authors are grateful to Asesh Krishna Datta for his careful reading
of the manuscript and many valuable suggestions. They are also thankful to
Debottom Das and Abhishek Iyer for discussion.

\providecommand{\href}[2]{#2}\begingroup\raggedright\endgroup


\end{document}